\documentclass[11pt, a4paper]{article}
\usepackage{jheppub}
\usepackage[utf8]{inputenc}
\usepackage[T1]{fontenc}
\usepackage{amsmath}
\usepackage{amssymb}
\usepackage{amsthm}
\usepackage{mathtools}
\usepackage{array}
\usepackage{enumerate}
\usepackage{physics}
\usepackage{pgfplots}
\usepackage{pgfplotstable}
\usepackage{tikz}
\usepackage{xcolor}
\usepackage{colortbl}
\usepackage{graphicx}
\usepackage{caption}
\usepackage{subcaption}
\usepackage{hhline}
\usepackage{array}
\usepackage{multirow}
\usepackage{standalone}
\usepackage{longtable}
\usepackage{xr}
\usepackage[noabbrev]{cleveref}
\usepackage{natbib}

\usetikzlibrary{matrix}
\usepgfplotslibrary{colormaps} 
\usepgfplotslibrary{colorbrewer}

\pgfplotsset{
    compat=newest}

\definecolor{myblue1}{RGB}{33,113,181}
\definecolor{myblue2}{RGB}{158,202,225}
\definecolor{mygreen1}{RGB}{49,163,84}
\definecolor{mygreen2}{RGB}{161,217,155}
\definecolor{cycle1}{RGB}{128,177,211}
\definecolor{cycle2}{RGB}{253,180,98}
\definecolor{cycle3}{RGB}{179,222,105}
\definecolor{cycle4}{RGB}{251,128,114}
\definecolor{cycle5}{RGB}{190,186,218}
\definecolor{cycle6}{RGB}{141,211,199}

\newcommand{\col}{\operatorname{col}}
\newcommand{\eff}{\operatorname{eff}}

\title{Large N Optimization for Multi-Matrix Systems}

\author[a,b]{Robert de Mello Koch,}
\author[c]{Antal Jevicki,}
\author[c]{Xianlong Liu,}
\author[b]{Kagiso Mathaba}
\author[b]{and Jo\~ao P. Rodrigues}

\affiliation[a]{School of Science, Huzhou University, \\
                Huzhou 313000, China}
\affiliation[b]{National Institute for Theoretical and Computational Sciences, \\
                School of Physics and Mandelstam Institute for Theoretical Physics, \\   
                University of the Witwatersrand, Wits, 2050, South Africa}
\affiliation[c]{Department of Physics, Brown University, \\
                182 Hope Street, Providence, RI 02912, USA}

\emailAdd{robert@neo.phys.wits.ac.za}
\emailAdd{antal\_jevicki@brown.edu}
\emailAdd{xianlong\_liu@brown.ed}
\emailAdd{0601228x@students.wits.ac.za}
\emailAdd{Joao.Rodrigues@wits.ac.za}

\abstract{
In this work we revisit the problem of solving multi-matrix systems through numerical large $N$ methods. The framework is a collective, loop space representation which provides a constrained optimization problem, addressed through master-field minimization. This scheme applies both to multi-matrix integrals ($c=0$ systems) and multi-matrix quantum mechanics ($c=1$ systems). The complete fluctuation spectrum is also computable in the above scheme, and is of immediate physical relevance in the later case. The complexity (and the growth of degrees of freedom) at large $N$ have stymied earlier attempts and in the present work we present significant improvements in this regard. The (constrained) minimization and spectrum calculations are easily achieved with close to $10^4$ variables, giving solution to Migdal-Makeenko, and collective field  equations. Considering the large number of dynamical (loop) variables and the extreme nonlinearity of the problem, high precision is obtained when confronted with solvable cases. Through numerical results presented, we prove that our scheme solves, by numerical loop space methods, the general two matrix model problem.
}

\begin{document}

\maketitle
\flushbottom

\section{Introduction} 

Large $N$ multi-matrix problems are at the center of many theories of current interest, involving membranes \cite{Hoppe:1986aj}, reduced super Yang-Mills theories \cite{Kazakov:2000aq, Corley:2001zk, Haggi-Mani:2000dxu, Aharony:2003sx, Ishibashi:1996xs}, field theory of critical and  noncritical strings \cite{Berenstein:2002jq, deMelloKoch:2002nq, Balthazar:2017mxh, Sen:2020eck, Das:1990kaa,Ishibashi:1993nqz, Jevicki:1993rr,Marino:2012zq,Gharibyan:2020bab,David:1984tx,Kazakov:1985ds,Kazakov:1985ea,Kazakov:1988ch,Gross:1990ay,Brezin:1989ss,Ginsparg:1990as}, phase transitions and black holes \cite{ Gross:1990md,Dalley:1992yy,Kazakov:2000pm,Maldacena:2005hi,Balthazar:2018qdv,Betzios:2017yms,Lowe:2017ehz,Amado:2017kgr,Cotler:2016fpe}, and M-atrix theory \cite{Anagnostopoulos:2007fw}. At the same time, apart from very special cases, these systems are not solvable due to the fact that they are highly nonlinear. At finite $N$ one has numerical Monte-Carlo methods \cite{Lucini:2003zr}, which have provided definite and most relevant results \cite{Anagnostopoulos:2007fw} with increase of simulations towards large $N$. The limiting theory, infinite $N$, features a rapid growth  of the degrees of freedom, represented by (Wilson) loop variables, which are physical, gauge invariant collective variables for the description of  matrix and non-Abelian gauge theories. As is well known, (Wilson) loops  become independent degrees of freedom at infinite $N$, and the exact theory is governed by non-linear Schwinger-Dyson (Migdal-Makeenko) equations \cite{Makeenko:1979pb}, or alternatively in the collective field theory representation, in terms of an effective action \cite{Jevicki:1980zg} and/or a collective Hamiltonian \cite{Jevicki:1979mb}. The later provides a unified approach in which $1/N=G$ appears as a coupling constant, and has been used through the years for non perturbative studies. For coupled systems (of matrices) at quantum level, only simple systems are solvable. However, numerical approaches were developed in \cite{one,Jevicki:1983wu}. In these previous studies the nature of the (infinite $N$) planar solution was understood as related to a constrained minimization problem, where a significant role is played by a set of inequalities associated with invariants (loops) in the collective description. An effective scheme for dealing with this constrained minimization was identified, through the use of master-field variables \cite{Jevicki:1983wu}. These methods were also seen to apply at sub-leading order in $N$ and in particular, to consideration of the spectrum \cite{Rodrigues:1983fs,Jevicki:1983hb}.
 
Recently, there is a renewed interest in large $N$ optimization \cite{Anderson:2016rcw}, with studies
\cite{Lin:2020mme,Han:2020bkb,Kazakov:2021lel} that overlap with earlier work, and which re-discover the importance of loop space inequality constraints.  
 
Due to potential high relevance in problems of emergent geometry, thermalization and black hole formation we revisit the earlier collective field constrained minimization and numerical master field methods, with interest in increasing the numbers of degrees of freedom, and the potential for high precision results. These are developed in the present work. For concreteness, and with purpose of making comparisons (with analytical results, when possible) we mostly deal with systems of two hermitian matrices, but the methods are seen to apply for any number, with arbitrary single or multi trace interactions.

The content of this paper is as follows: In the overview Section \ref{sec:overview} we provide a short summary of the collective large $N$ method \cite{Jevicki:1983wu}. In particular, the infinite $N$ constrained minimization scheme is summarized, featuring the associated complete set of (loop space) inequalities. We then give a summary of the unconstrained master variable method which applies for the large $N$ ground state and also for spectrum studies \cite{Rodrigues:1983fs,Jevicki:1983hb}. We explain the study of multi-matrix integrals through the Hamiltonian description \cite{Rodrigues:1985aq} given by Fokker-Planck. In Section \ref{sec:methods} we describe the numerical methods used for constrained minimization. In Section \ref{sec:SD-models} we give a list of models studied and present the numerical solution of the associated matrix integrals. In Section \ref{sec:matrix-quantum-mechanics-and-spectrum} we present results, giving numerical solutions for one- and two-matrix large $N$ quantum Hamiltonian problems. At the large $N$ level these essentially correspond to giving a (numerical) solution of the fully nonlinear Schwinger-Dyson equations with approximately $10^4$ loop variables. Correspondingly methods (and evaluation) of large $N$ ground state energies, gaps and low lying spectra are also given. Conclusions and future applications to problems of interest are commented in Section \ref{sec:conclusions}.

\section{Overview}
\label{sec:overview}

In this article we will study and develop numerical techniques for solving the large $N$ multi-matrix theories. We will consider two classes of the multi-matrix systems. The first one is the multi-matrix integral ($c=0$ systems) whose partition function reads
\begin{equation}
    \int \prod_{l=1}^{d} \dd M_{l} \: e^{-S(M_1, M_2, \dots, M_d)} \, ,
\end{equation}
where $S(M)$ is a multi trace action. The second one is matrix quantum mechanics (MQM) ($c=1$ systems), whose dynamics is given by a Hamiltonian of the form
\begin{eqnarray} \label{eq:MQM_Hamiltonian}
H=\frac{1}{2}{\rm Tr}(\Pi_1^2+\Pi_2^2+\dots+\Pi_d^2)+{V(M_1,M_2,\dots,M_d)} \, ,
\end{eqnarray} 
where $\Pi_l$ is the canonical conjugate of $M_l$. 

The large $N$ expansion in the \emph{collective field formulation} is developed after a change of variables from the original matrix valued variables to invariant variables, which we refer to as ``loops''. This terminology has its origins in lattice gauge theory, where the basic field degrees of freedom are unitary matrices $U_l$, one for each link in the lattice. In that case the invariant variables, Wilson loops $\phi(C)$, are obtained by taking the trace of an ordered product of unitary matrices, one for each link of a closed path $C$. It is common to continue to refer to invariant variables as ``loops'' even for theories of hermitian matrices $M_l$ where invariants are given by the trace of products of the matrices. As an example, for the case of two hermitian matrices we have
\begin{equation} 
\phi (C)=\operatorname{Tr}(M_1^{n_1}M_2^{n_2}M_1^{n'_1}M_2^{n_2'}\cdots) \, .
\end{equation}
In this case the invariant $\phi(C)$ is not labeled by a closed path, but rather by specifying a word $C$ in the alphabet of the matrices. The word specifies the order in which matrices are multiplied before tracing. The invariant loop variables are then described by all of the words with cycling identification. For example, $\Tr(M_1 M_1 M_2)$ is equal to $\Tr(M_1 M_2 M_1)$ and $\Tr(M_2 M_1 M_1)$ due to the cyclicity of trace, hence they all refer to the same invariant loop variable. It is in this sense that we use the loop terminology. For the purpose of counting the number of loops, it is useful to enumerate the loops with a permutation $\sigma$. The loop corresponding to permutation $\sigma$ is denoted $\phi_\sigma$. Consider loops built as single trace products of $n$ $M_1$s and $m$ $M_2$s. The permutation $\sigma\in S_{n+m}$ specifies the loop $\phi_{\sigma}$ as follows
\begin{align}
\phi_{\sigma} = & {\rm Tr} (\sigma M_1^{\otimes n}M_2^{\otimes m}) \nonumber \\
 = & (M_1)_{i_1\, i_{\sigma(1)}}\cdots (M_1)_{i_n\, i_{\sigma(n)}}
(M_2)_{i_{n+1}\, i_{\sigma(n+1)}}\cdots (M_2)_{i_{n+m}\, i_{\sigma(n+m)}} \, .
\end{align}
The cycle structure of the permutation translates into the trace structure of $\phi_{\sigma}$. For a single trace $\sigma$ must be an $n+m$ cycle. This parametrization is not unique as distinct permutations do not necessarily define distinct loops. This follows by noting that
\begin{align}
\phi_{\tau^{-1}\sigma\tau} = & {\rm Tr} (\sigma\tau M_1^{\otimes n}M_2^{\otimes m}\tau^{-1}) \nonumber \\ 
 = & (M_1)_{i_{\tau(1)}\, i_{\sigma(\tau(1))}}\cdots (M_1)_{i_{\tau(n)}\, i_{\sigma(\tau(n))}}
(M_2)_{i_{\tau(n+1)}\, i_{\sigma(\tau(n+1))}}\cdots (M_2)_{i_{\tau(n+m)}\, i_{\sigma(\tau(n+m))}} \, ,
\end{align}
so that if $\tau\in S_n\times S_m=$ the permutation group swapping $M_1$s with each other and $M_2$s with each other, then
\begin{eqnarray}
\phi_{\tau^{-1}\sigma\tau}=\phi_{\sigma}\label{redundant} \, .
\end{eqnarray}
In general, two loops $\phi_{\sigma_1}$ and $\phi_{\sigma_2}$ are the same if
\begin{eqnarray}
\tau^{-1}\sigma_1\tau =\sigma_2\label{loopEquality}
\end{eqnarray}
for some $\tau\in S_n\times S_m$. Removing this redundancy, we are left with a complete set of (infinitely many) loops. To develop some intuition for this description, note that if $\tau$ belongs to the cyclic group $\mathbb{Z}_{n+m}$ generated by the $n+m$ cycle given by $(123\cdots n+m)$, then the equality (\ref{redundant}) expresses nothing but the cyclicity of the trace. In general, we need to divide by more than just cyclicity and (\ref{redundant}) is a convenient way to correctly account for all redundancies.

A relevant class of observables for the $c = 0$ systems are provided by correlation functions of the loops
\begin{eqnarray}
\langle\phi (C)\rangle &=& \Big\langle\Tr \big(M_1^k M_2^l M_3^m\cdots\big)\Big\rangle\cr
&=&\int \prod_{a,b=1}^N \prod_{l=1}^d \dd (M_l)_{ab}\,\, e^{-S}\, \Tr \big(M_1^k M_2^l M_3^m\cdots\big)\, .
\label{loopvev}
\end{eqnarray}
These expectation values can be determined through equations of motion for the loop expectation values. The loop equation is derived as a Schwinger-Dyson equation for the loops. In the case of a theory of unitary matrices the equation of motion for the Wilson loops are known as the Migdal-Makeenko loop equations \cite{Makeenko:1979pb}. For the case of hermitian matrices, the loop equations follow by inserting the matrix derivative under the integral: 
\begin{equation} \label{eq:SD_eq}
    0=\int \prod_{l=1}^{d} \dd M_l \:
    \sum_{a=1}^{d} \pdv{}{(M_a)_{ij}}
    \left( \pdv{\phi(C)}{(M_a)_{ji}}e^{-S}\right) \, .
\end{equation}
The loop equation is a quadratic equation in the large $N$ limit which takes the form
\begin{equation}
\sum_{C_1,C_2}p(C;C_1,C_2)\langle\phi(C_1)\rangle \langle\phi(C_2)\rangle
-\sum_s j(C,s;C')\bigg\langle\phi(C')\frac{\partial S}{\partial\phi(s)}\bigg\rangle = 0 \, .
\end{equation}
The integer $p(C;C_1,C_2)$ specifies the number of ways in which a loop $C$ can be split into loops $C_1$ and $C_2$, while the integer $j(C,s;C')$ specifies the number of ways loops $C$ and $s$ can be joined to produce $C'$. Both will be described in detail below. Solving these non-linear coupled equations is highly non-trivial. Using collective field theory one obtains an effective potential which, when minimized, gives the large $N$ solution to the Schwinger-Dyson equations.

As for the $c = 1$ MQM systems, changing to invariant variables we obtain a collective potential $V_{\col}$, which when minimized, again determines the large $N$ expectation values of loops. These will be discussed in detail below.

\subsection{Collective (Loop Space) Representation}

The collective representation of multi-matrix systems can be given both at the action level for matrix integrals ($c=0$), and at the Hamiltonian level for coupled quantum mechanical ($c=1$) systems. These are closely related, so in formulating numerical methods one can work in the Hamiltonian framework \cite{Jevicki:1979mb}. To proceed, let us define the adjoint of the loop variable. For a loop $\phi(C)$, we define its adjoint as $\phi(\bar{C})$, where $\bar{C}$ is the reverse of the word $C$. For example, for a word $C = abaab$, where $a = M_1$ and $b = M_2$, its reverse is $\bar{C} = baaba$. In hermitian matrix models, this corresponds to taking a hermitian conjugate of the matrix products. As such the adjoint $\phi(\bar{C})$ is simply the complex conjugate of $\phi(C)$ in hermitian matrix systems:
\begin{equation}
    \phi(\bar{C}) = \bar{\phi}(C) \, .
\end{equation}
The ``bar'' symbol on the right hand side of the above formula denotes complex conjugate. We note in the hermitian one-matrix models case, the adjoint of a loop is itself since all loops are real valued. The use the loop adjoints makes the hermiticity of the collective representation manifest as we will see below.

\paragraph{Matrix integrals.} 
Let us start with the collective representation of the multi-matrix integral problem with any number $d$ of matrices which at large $N$ is efficiently described using an effective action. This is obtained by a change of integration variables (from matrices to loops)
\begin{equation}
\int \prod_{a,b=1}^N \prod_{l=1}^d \dd (M_l)_{ab}\,\, e^{-S} 
= \int \prod_C \dd \phi(C)\, J (\phi) \, e^{-S}
= \int \prod_C \dd \phi(C)\, e^{-S_{\rm eff}} \, ,
\end{equation}
resulting in a large $N$ collective action $S_{\rm eff} = S -\ln \, J$. The  main ingredient in this effective description is the Jacobian $J$ and its form is in general specified by the collective formalism. Variation of this collective action correctly produces the Schwinger-Dyson (SD) equations \eqref{eq:SD_eq}. The derivative of the collective action \cite{Jevicki:1980zg} gives 
\begin{eqnarray} \label{eq:SD}
\bar{\omega}(C)-\sum_{C'}\Omega(C,C')\pdv{S}{\phi(C')} = 0 \, .
\end{eqnarray}
The functional $\omega(C)$ stands for
\begin{equation}
    \omega (C; \phi ) 
    \equiv 2 \hat E_l^\alpha \hat E_l^\alpha \phi (C) 
    = \sum_{(C_{1}, C_{2})} p (C; C_1 , C_2 ) \, \phi (C_1 )\, \phi (C_2) \, ,
\end{equation}
representing the splitting of contour $C$ into sub-contours $(C_1 , C_2 )$.\footnote{
    For unitary $(M_l)_{ab}$ matrices,  $\hat{E}_l^{\alpha}=\sqrt{2}\,t_{ab}^{\alpha}(M_l)_{bc}\,\frac{\partial}{\partial(M_l)_{ac}}$. The generators of the Lie algebra of $U(N)$ are normalized such that $\sum_\alpha  t_{ab}^{\alpha}t_{a'b'}^{\alpha}=\frac{1}{2}\delta_{ab'}\delta _{a'b}$. For hermitian matrix systems, $\hat{E}_l^{\alpha}\to\hat{E}_l^{ab}=-i\frac{\partial}{\partial (M_l)_{ba}}$, with $(M_l)_{ab}$ hermitian.}
The split occurs at the pinched link, so in general there are several distinct ways for the process, which necessitates the sum. The integer $p(C;C_1,C_2)$ counts the number of ways loop $C$ can be partitioned, by the splitting operation, into loops $C_1$ and $C_2$. In a similar way one has
\begin{equation}\label{SymOmeg}
    \Omega (C_1,C_2; \phi ) 
    \equiv -2 \hat E_l^\alpha \, \bar{\phi} (C_1) \, \hat E_l^\alpha \, \phi (C_2) 
    = \sum_{C} \, j(C_1,C_2;C) \, \phi (C)
\end{equation}
for the opposite operation of joining contours. We note the use of loop adjoint in the definition. The integer $j(C_1,C_2;C)$ counts the number of ways in which $C_1$ and $C_2$ can be joined to produce $C$. A computer algorithm was developed to generate loops and also these loop processes which we call \emph{loop algebra}.

For numerical minimization it is useful to follow \cite{Rodrigues:1985aq}, where it was established (through a stochastic quantization) that the above SD problem can be represented through (Fokker-Planck type) the quantum mechanical Hamiltonian: $H =K + V_{\rm eff}$. The solution of nonlinear SD equations is then obtained by minimization of an effective potential of the following form
\begin{align}
V_{\rm eff} 
  = \frac{1}{8} \bar{\omega}(C) \Omega^{-1}(C,C') \,\omega(C')
   -\frac{1}{4} \bar{\omega}(C)\pdv{V}{\phi(C)}
   +\frac{1}{8}\overline{\pdv{S}{\phi(C)}}\Omega(C,C')\pdv{S}{\phi(C')} \, , \label{MPPot}
\end{align}
where the ``bar'' symbols again denote complex conjugates. This effective potential also gives the leading large $N$ configuration of the bosonic sector of multi matrix Marinari-Parisi \cite{Marinari:1990jc} type models. Expansion around the stationary point leads to equations for small fluctuations and a systematic $1/N$ expansion scheme.

\paragraph{Matrix quantum mechanics.}
We now describe the collective field formulation of the large $N$ MQM in detail. Considering a general multi-matrix quantum mechanics problem, in the operator formalism, one has a transition to the collective description by performing a change to curvilinear (loop space) variables (with a Jacobian $J$) which induces the collective Hamiltonian, taking the form
\begin{equation}\label{Ham_S_1}
    H_{\rm col} = \frac{1}{2} \sum_{C, C'} \, \pi^{\dagger }(C) \, \Omega(C,C') \pi (C') + V_{\rm col} [\phi ] \, 
\end{equation}
with
\begin{equation} \label{Ham_S_2}
    V_{\rm col}[\phi] =  \frac{\hbar}{8} \, \sum_{C, C'} \bar{\omega} (C) {\Omega}^{-1} (C,C') \, \omega(C') + V[\phi] \, ,
\end{equation}
and $\pi (C)$ representing the conjugates to the loops $\phi (C)$. Here $V[\phi]$ is the original potential written in terms of loops. The collective Hamiltonian is manifestly hermitian, and is equivalent to the original MQM Hamiltonian \eqref{eq:MQM_Hamiltonian}. We have used the notation $V_{\rm col}$ to denote the \emph{collective potential} obtained for MQM systems, which plays an analogue role of the effective potential $V_{\rm eff}$ for matrix integral problems, as described above.

\paragraph{Notation emphasis.} To proceed, let us emphasize that throughout we will use $V_{\eff}$ to denote the \emph{effective potential} associated with matrix integrals ($c=0$ systems), and $V_{\col}$ to denote the \emph{collective potential} associated with MQM ($c = 1$ systems).

\paragraph{Large $N$ background and fluctuation spectrum.}
In the next subsection we explain that the $N\to\infty$ limit is obtained by minimizing the potential $V_{\eff}$ or $V_{\col}$ and that this minimization is subject to a sequence of inequalities which constrain the range of the loop variables. The relevance of constrained minimization becomes even more fundamental when one proceeds to study the spectrum at large $N$. Ordinarily this would be given directly in loop space by expanding about the stationary field
\begin{align}
\phi (C) & = \phi_0 (C) + \frac{1}{N} \eta (C), \\
\pi (C) & = NP(C),
\end{align}
and reducing the collective Hamiltonian to a quadratic, small fluctuation Hamiltonian
\begin{equation}
H^{(2)}_{\col} = \frac{N^2}{2}\sum_{C,C'} \, P^{\dagger} (C) \Omega^0 (C,C^{\prime})P (C^{\prime})
+\frac{1}{2 N^2}\sum_{C,C'} 
\bar{\eta} (C) V^{(2)}(C,C^{\prime})\eta (C^{\prime})\, ,
\end{equation}
\begin{equation}
\Omega^0 (C,C^{\prime})=\Omega (C,C^{\prime})\Big|_{\phi_0(C)} \, ,\qquad
V^{(2)}(C,C^{\prime}) = \left.\pdv[2]{V_{\rm col}}{\bar{\phi}(C)}{\phi (C^{\prime})} \right|_{\phi_0(C)} \, .
\end{equation}
This $H^{(2)}$, and its diagonalization provides the spectrum at large $N$. Essentially, it is determined by eigenvalues of the matrix:
\begin{equation}\label{LoopSpectrum}
\varepsilon_i^2  =  {\rm eig }(\Omega^0 V^{(2)}) = {\rm eig} (V^{(2)}\Omega^0)\, .
\end{equation}
As we will explain next, these are to be found subject to obeying a set of (loop) space inequalities that are central for reaching the correct minima.

\subsection{Loop Space Inequalities and Constrained Minimization} 

\paragraph{Positivity of the loop joining matrix $\Omega$.} 
In the collective Hamiltonian description, the $N \rightarrow \infty$ limit (and the sum of planar diagrams) is given by the semiclassical approximation. The problem is therefore to solve for the static stationary configuration denoted $\phi_0 (C)$, which minimizes the potential $V_{\rm col}$. This would be generally given by the equation
\begin{equation}
\pdv{V_{\rm col}[\phi]}{\phi (C)} = 0\, .
\end{equation}
However, as was understood earlier \cite{one,Jevicki:1983wu}, in the minimization a role is also played by a sequence of inequalities (analogous to Schwarz inequalities) which constrain the range of loop variables. Such a sequence of inequalities is generic, and will be present for any set of variables representing invariants. In the collective description the inequalities are directly visible and can be generally specified and given in terms of the loop space matrix~$\Omega$. This matrix participates in the kinetic term of the Hamiltonian and defines the loop space symplectic form. As such the matrix $\Omega$ must be positive semi-definite. Indeed from the definition of the matrix $\Omega$ one has the fact that it can be written as 
\begin{equation}
\Omega \left( C, C^{\prime} \right) = \sum_i  \,  \bar{A}_{i C}A_{i C^{\prime}} = \sum_i  \,  A^\dagger_{C i} A_{i C^{\prime}} \, ,
\end{equation}
i.e. it is explicitly \emph{positive semi-definite}. Indeed, in our case, defining\footnote{Here we consider the case that $(M_l)_{ab}$ is a unitary matrix, but the same conclusion holds when $(M_l)_{ab}$ is hermitian.}
\begin{equation}
A_{i C} \equiv \pdv{\phi (C)}{(M_l)_{ab}} \, , \quad\quad i \equiv (a, b, l) \, ,
\end{equation}
we obtain
\begin{align*}
\Omega (C, C^{\prime}) & 
= - 2 \sum_{\alpha} \hat{E}^{\alpha} \bar{\phi} (C) \hat{E}^{\alpha} \phi (C^{\prime}) \\
& =  2 \sum_{l \alpha} (M^{\dagger}_l)_{ca}t_{ab}^{\alpha} \, \pdv{\bar{\phi}(C)}{(M^{\dagger}_l)_{cb}}\, t_{a'b'}^{\alpha} (M_l)_{b'c'} \, \pdv{\phi (C^{\prime} )}{(M_l)_{a'c'}}\\
& =   \sum_{l} \pdv{\bar{\phi}(C)}{(M^{\dagger}_l)_{cb}}\, \pdv{\phi (C^{\prime} )}{(M_l)_{bc}}
= \sum_{l} \pdv{\bar{\phi}(C)}{(\bar{M}_l)_{bc}}\, \pdv{\phi (C^{\prime} )}{(M_l)_{bc}} \\
& =\sum _i \bar{A}_{i C} A_{i C^{\prime}} \, .
\end{align*}

\paragraph{Positivity constraints.}
\emph{The minimization therefore must be done subject to the positivity condition of the loop joining matrix $\Omega$. In general the positive semi-definiteness of the loop valued matrix $\Omega$ gives the complete set of inequalities, which schematically reads} 
\begin{equation}
    \operatorname{eig}\big(\Omega(C, C')\big) \geq 0.
\end{equation}
We can write the complete set of these generalized loop space (Schwarz) inequalities as follows. A convenient basis (and explicit form) turns out to be given by the sequence of the sub-determinants $\det_k (\Omega)\ge 0$. The sub-determinant is defined by (repeated indices are summed)
\begin{eqnarray}
\operatorname{det}_l (\Omega) &=& \frac{1}{l!(N_\Omega -l)!}
\epsilon_{C_1\cdots C_l a_1\cdots a_{N_\Omega-l}}
\epsilon_{C_1\cdots C_l a_1\cdots a_{N_\Omega-l}}
\Omega (C_1 ,C_1')\cdots \Omega (C_l ,C_l')\cr
&=&\frac{1}{l!(N_\Omega -l)!}
\epsilon_{C_1\cdots C_l a_1\cdots a_{N_\Omega-l}}
\epsilon_{C_1'\cdots C_l' a_1\cdots a_{N_\Omega-l}}
\bar{A}_{i_1 C_1}A_{i_1 C_1'}\cdots \bar{A}_{i_l C_l}A_{i_l C_l'}\cr
&=& \bar{T}_{i_1\cdots i_l a_1\cdots a_{N_\Omega-l}} T_{i_1\cdots i_l a_1\cdots a_{N_\Omega-l}} \, ,\label{subdet}
\end{eqnarray}
where $N_{\Omega}$ is the dimension of the loop space joining matrix $\Omega$ and
\begin{equation}
T_{i_1\cdots i_l a_1\cdots a_{N_\Omega-l}}
=\frac{1}{\sqrt{l!(N_\Omega -l)!}}
\epsilon_{C_1'\cdots C_l' a_1\cdots a_{N_\Omega-l}}
A_{i_1 C_1'}\cdots A_{i_l C_l'} \, .
\end{equation}
The expression in the last line of \eqref{subdet} makes the positivity of the sub-determinant manifest. An alternative formula for the sub-determinant is provided by
\begin{equation}
\operatorname{det}_l (\Omega) =\chi_{(1^l)}(\Omega) \, ,
\end{equation}
where $\chi_{(1^l)}(\cdot)$ is a Schur polynomial and $(1^l)$ is the Young diagram with a single column and $l$ rows. This formula is simple with the normalization chosen in \eqref{subdet}. The sub-determinant basis is particularly useful when some of the constraints are saturated, in which case certain eigenvalues of $\Omega$ vanish. If $p$ eigenvalues vanish, then there are $p$ independent vanishing sub-determinants
\begin{equation}
\operatorname{det}_{k}(\Omega)=0 \, , \qquad k=N_{\Omega}-p+1,N_{\Omega}-p+2, \dots, N_{\Omega} \, .
\end{equation}

Once expressed in terms of loops, this positivity condition implies highly non-trivial inequality constraints among the $\phi (C)$'s. The positivity inequalities constrain the eigenvalues of the loop space matrix $\Omega$, leading to a constrained minimization of $V_{\rm eff}$ or $V_{\col}$.\footnote{
    Numerically and for unitary matrices systems, one finds that down to a certain critical value of the coupling, a standard unconstrained minimization procedure converges giving the correct minima. However, at a critical point (and below) the procedure breaks down. For single unitary matrix systems this is the Gross-Witten phase transition \cite{Gross:1980he}, also present in the single unitary matrix hamiltonian systems \cite{JSW,Rodrigues:1981sd,Rodrigues:1982qr}.}
The stationary minimum solution is generally characterized by saturation of a certain number of inequalities, whereby the loop space matrix $\Omega$ develops zero eigenvalues. 

\subsection{Master Field}

A most complete way to deal with the constrained optimization at large $N$ is through variables that \emph{automatically} assure the positivity condition of the loop space matrix $\Omega$. These are the \emph{master field variables}, or simply the original matrix valued variables of the system. In terms of such variables the expectation values of loops are determined by a direct stationary-point equation of the loop space effective ($c=0$) or collective ($c=1$) potential
\begin{equation}
\pdv{}{(M_l)_{ab}} \, V_{\rm eff/col} = 0\, .
\end{equation}
Here the notation $V_{\rm eff/col}$ represents either $V_{\eff}$ or $V_{\col}$. This saddle-point equation in terms of the master variables is correct in all phases of the theory, both the weak- and strong-coupling. Let us also mention how the correct loop-space equation follows from the master equation. Multiplying and summing over the appropriate factors, we obtain
\begin{equation}
\sum_{ab,l}\,\sum_{C^{\prime}}\, \pdv{\bar{\phi}(C)}{(\bar{M}_l)_{ab}}\,
\pdv{\phi(C^{\prime})}{(M_l)_{ab}}
\, \pdv{V_{\rm eff / col}}{\phi (C')}=\sum_{C'}{\Omega}(C,C^{\prime}) \, 
\pdv{V_{\rm eff/col}}{\phi (C')} = 0 \, ,
\end{equation}
This fact that the saddle-point equation in terms of the master variable produces the correct loop-space equation can be taken as another argument for the existence of the master field. 

Denote the set of master variables and their complex conjugates by $\phi_{\alpha}$ and $\bar{\phi}_{\alpha}$, respectively. To see that the positivity of $\Omega$ is assured when working in terms of the master variables, note that we can write $\Omega$ as
\begin{align}
\Omega (C, C^{\prime} ) & = \sum_{\alpha} \, \pdv{\bar{\phi}(C)}{\bar{\phi}_{\alpha}} \, \pdv{\phi (C^{\prime} )}{\phi_{\alpha}} = \sum_{\alpha} \, A^{\dagger}_{C\alpha} \, A_{\alpha C'} ,\qquad 
A_{\alpha C} = \pdv{\phi (C)}{\phi_{\alpha}} \, . \label{OmegaDef}
\end{align}
The set $\{\phi_{\alpha} , \bar{\phi}_{\alpha}\}$ is always assumed to be at least as large as the set of invariants $\{\phi (C)\}$. We can think of these variables as the original variables of the theory, in which case they transform non-trivially under the existing internal symmetries and form a larger set than that of the invariants that can be obtained from them. Nevertheless, we have also other situations in mind: one may truncate the effective potential, in which case the number of $\phi_{\alpha}$ variables must be larger than the number of loop variables included in the potential. In general, the set $\{\phi_{\alpha}\}$ is at least as large as the set of invariants. This is the case in one-matrix models, for instance, where one identifies the $\phi_{\alpha}$'s with the matrix eigenvalues.  

The important characteristic of these variables then is that they solve the positivity constraint explicitly. Therefore, in the large-$N$ limit, the ground state configuration is determined by the saddle-point condition
\begin{equation}\label{saddle}
\pdv{V_{\rm eff/col}}{\phi _{\alpha} } = \sum_C \, \pdv{\phi (C)}{\phi_{\alpha} }\, \pdv{V_{\rm eff/col}}{\phi (C)} = 0 \, .
\end{equation}

It is clear that it is useful to reduce the problem to unconstrained minimization.  In \cite{one,Jevicki:1983wu} we have presented and tested numerically such a framework developing a hybrid loop space $+$ master field approach. The use of master variables \cite{Jevicki:1983wu} is simple. In terms of the original variables the inequalities are {\it automatically} obeyed so we can think of changing back to these master variables but keeping the collective loop space Hamiltonian. This is because it is the loop space Hamiltonian that generates the $1/N$ expansion. Now the ground state and the fluctuation spectrum will be given by unconstrained minimization, satisfying 
\begin{equation}
\pdv{V_{\rm eff/col} [ \phi (C\{\phi_\alpha\}) ]}{\phi_{\alpha} }= 0 \, , \qquad \forall \phi_{\alpha} \, .
\end{equation}
This represents an implicit equation for the master field $\phi_{\alpha}$ since it enters $V_{\rm eff/col}$ through the loop variables 
$\phi (C\{\phi_\alpha\})$. As such it is not very useful at the analytic level but it can be easily implemented numerically \cite{one}.

We will consider systems of two hermitian matrices in this article.  Systems of hermitian matrices are always in the phase in which some of the constraints are saturated. Consequently for these systems the use of master variables is of particular importance. The loop invariants consist of single trace products of these matrices. As a result, one of the matrices can be chosen to be diagonal, and the other will be parametrized in the Lie algebra of the unitary group. The master variables are then real, which is something we use in the following.\footnote{
    The general case is discussed in \cite{Jevicki:1983hb}.} 

The master eigenvalue equations for the fluctuation spectrum are similarly obtained in this hybrid scheme. Denote in our Hamiltonian formulation the master field solution by $\phi^0_\alpha$. 
We are then led to appropriate spectrum eigenvalue equations through a shift in the loop space collective Hamiltonian. One essentially considers a `canonical' transformation \cite{Jevicki:1983hb} from loops to master fields
\begin{align}  
P_{\beta} = \sum_C \, \pdv{\phi (C)}{\phi_{\beta}} \bigg\vert_{\phi^0_\alpha} P(C), 
\qquad
\eta_{\beta} = \sum_{C} \, \left[\pdv{\phi (C)}{\phi_{\beta}} \bigg\vert_{\phi^0_\alpha}\right]^{-1} \eta(C) \, ,
\end{align}
to obtain the following quadratic Hamiltonian
\begin{equation}
H^{(2)} = \frac{1}{2} \sum_{\alpha} \, P_{\alpha} P_{\alpha} 
+ \frac{1}{2} \sum_{\alpha, \beta} \, {\eta}_{\alpha} \, \mathcal{M}_{\alpha \beta} \, {\eta}_{\beta} \, .
\end{equation}
Note that the term linear in $\eta$ vanishes as a result of equation (\ref{saddle}). The mass matrix $\mathcal{M}_{\alpha \beta}$ is then to be computed from
\begin{equation} \label{MasterSpectrum}
\mathcal{M}_{\alpha\beta} =  \pdv{\bar{\phi} (C)}{\phi_{\alpha}} \bigg\vert_{\phi^0_\alpha}
\left.\pdv[2]{V_{\col}}{\bar{\phi} (C)}{\phi (C^{\prime})} \right|_{\phi_0\{\phi^0_\alpha \}} 
\pdv{\phi (C')}{\phi_{\beta}} \bigg\vert_{\phi^0_\alpha} \, .
\end{equation}
Here the repeated indices $C$ and $C^{\prime}$ are summed, which in principle range from 1 to infinity. To compute it numerically we must perform the truncation discussed below. The fluctuation spectrum is obtained by solving for the square root of the nonzero eigenvalues of the mass matrix. In Section \ref{sec:matrix-quantum-mechanics-and-spectrum} we will show in detail that this is equivalent to the direct computation of fluctuation spectrum in loop space.

We will see that in numerical evaluations of multi-matrix problems, the hybrid loop space $+$ master field formalism turns out to be advantageous, converging rapidly and giving excellent agreement already at small loop and color cutoffs. With interest in precision optimization we will furthermore test the method for larger sizes and number of minimization variables ($\sim 9 \times 10^3$).

\section{Methods}
\label{sec:methods}

\subsection{Loop truncation}
\label{subsec:loop_trucation}

A numerical implementation of the large $N$ minimization of the collective potential necessarily involves a truncation of the infinite dimensional loop space. We will truncate the loop space by restricting to loops that limit the number of matrices appearing in the trace to be smaller than some fixed cut off $L_{\mathrm{max}}$. This loop truncation is reminiscent of level truncation in string field theory \cite{Kostelecky:1988ta}. To gather some insight into this truncation, in this section we consider the counting of loops as a function of the cut off. A convenient approach towards this counting uses permutations to enumerate loops.

\paragraph{Loop counting.} 
Consider the hermitian two-matrix model. The truncation of loop space that we employ restricts $n+m\le L_{\rm max}$ a fixed maximum loop length, where $n$ and $m$ denote the number of $M_1$ and $M_2$ in a loop, respectively. It is interesting to count the number of loops as a function of $L_{\rm max}$. This counting is a useful input to the numerical implementation since it specifies how many variables appear in the numerical minimization. There is an extremely rapid growth of the number of loops with increasing $L_{\rm max}$. The counting problem can be solved with a standard application of Polya theory \cite{Bianchi:2003wx}. We start by introducing the single letter partition function, which for two matrices is
\begin{equation}
Z_1=x+y\,.
\end{equation}
The single trace partition function, which counts the number of loops, is now given by
\begin{eqnarray}
F(x,y)&=&\sum_{n}\sum_{n|d}\frac{\varphi (d)}{n}Z_1 (x^d,y^d)^{\frac{n}{d}}\cr
&=&\sum_{n,m=1}^\infty {\cal N}_{n,m}x^n y^m \,.
\end{eqnarray}
The sum is over all integers $n$. At each $n$ there is a second sum over $d$ which runs over the divisors of $n$, i.e. all the integers that can be divided into $n$ without remainder. The function $\varphi (d)$ is the Euler totient function. 
The degree $L$ contribution to $F(x,y)$ counts single trace operators constructed from $L$ matrices. The coefficient ${\cal N}_{n,m}$ of the monomial of degree $n$ in $x$ and degree $m$ in $y$ counts the number of loops that can be constructed using $n$ $M_1$s and $m$ $M_2$s. The first few terms of $F(x,y)$ are
\begin{equation}
\begin{split}
F(x,y) = & (x+y)+\left(x^2+x y+y^2\right)
+\left(x^3+x^2 y+x y^2+y^3\right) \\
& +\left(x^4+x^3 y+2 x^2 y^2+x y^3+y^4\right) \\
& +\left(x^5+x^4 y+2 x^3 y^2+2 x^2 y^3+x y^4+y^5\right) \\
& +\left(x^6+x^5 y+3 x^4 y^2+4 x^3 y^3+3 x^2 y^4+x y^5+y^6\right)
+\cdots
\end{split}
\end{equation}
To interpret this answer, note that for example, the contribution $2x^3 y^2$ implies that there are two independent loops that can be constructed using 3$M_1$s and 2$M_2$s. These two loops are ${\rm Tr}(M_1^3 M_2^2)$ and ${\rm Tr} (M_1^2 M_2 M_1 M_2)$.

To explore how the total number of loops grows, it is useful to consider the blind partition function, obtained by setting $y=\alpha=x$. The coefficient of $\alpha^n$ counts the total number of loops constructed using $n$ matrices. The blind partition function is
\begin{equation}
\begin{split}
F(\alpha,\alpha)=&
2 \alpha +3 \alpha ^2+4 \alpha ^3+6 \alpha ^4+8 \alpha ^5+14 \alpha ^6+20 \alpha ^7
+36 \alpha ^8+60 \alpha ^9+108 \alpha ^{10}+188 \alpha ^{11} \\
&+ 352 \alpha ^{12}+632 \alpha ^{13}+1182 \alpha ^{14}+2192 \alpha ^{15}
+4116 \alpha ^{16}+7712 \alpha ^{17}+
+14602 \alpha ^{18} \\
&+ 27596 \alpha^{19}+52488 \alpha^{20}+99880 \alpha^{21}+190746 \alpha^{22}
+364724 \alpha ^{23}+699252 \alpha ^{24} \\
&+ 1342184 \alpha ^{25}+2581428 \alpha ^{26}+4971068 \alpha ^{27}+9587580 \alpha ^{28}+\cdots
\end{split}
\end{equation}
demonstrating an extremely rapid growth in the number of invariants (loops).

\paragraph{Loop space truncation.}
Our numerical implementation of loop space dynamics truncates to the subspace of invariants, given by all loops with $L_{\rm max}=2l-2$ matrices or less in the trace. In this scheme $\Omega$ is an $N_\Omega\times N_\Omega$ matrix, where $N_\Omega$ is the number of loops with $l$ matrices or less in the trace. $\Omega$ itself depends on a total of $N_{\rm Loops}$, which is the number of loops with $2l-2$ matrices or less. For this reason, our minimization scheme minimizes with respect to $N_{\rm Loops}$ independent variables. The values of $L_{\rm max}$, $N_\Omega$ and $N_{\rm Loops}$ for values $3\le l\le 10$ are given in Table \ref{table:1} below.

\begin{table}[h]
\begin{center}
\begin{tabular}{||c|c|c||} 
\hline
$L_{\rm max}$  & $N_\Omega$  & $N_{\rm Loops}$ \\ [0.5ex] 
\hline\hline
4 & 9 & 15\\ 
\hline
6 & 15 &37\\ 
\hline
8 & 23 &93\\ 
\hline
10 & 37 &261\\ 
\hline
12 & 57 &801\\
\hline
14 & 93 &2615\\
\hline
16 & 153 &8923\\
\hline
18 & 261 &31237\\ [1ex] 
 \hline
\end{tabular}
\caption{Truncating loop space.}
\label{table:1}
\end{center}
\end{table}
For the minimization, we parametrize $M_1$ and $M_2$ using $N+N^2$ real valued master variables. These are the $N$ eigenvalues for $M_1$ and the $N^2$ matrix elements in $M_2$. For a given $l$ we must choose the number of colors $N$ large enough that $N^2+N\ge N_{\rm Loops}$. To obtain the large $N$ background, our numerical experiments focus on $l=9$, so that we keep a total of 8923 loops. We choose $N=94$ so that there are a total of 8930 master variables.

\paragraph{Spectrum calculation.}
To obtain the fluctuation spectrum based on master variables, one needs to preserve the matrix structure of the loop derivatives \eqref{MatDeriv} with respect to both $M_1$ and $M_2$. We then truncate the mass matrix equation \eqref{MasterSpectrum} as follows
\begin{equation}\label{MasterSpectrumTrunc}
\mathcal{M}_{ij} =  \sum_{C,C' = 1}^{N_{\text{Loops}}}
\bar{A}_{iC}\Big|_{\phi_\alpha^0}
\left.
    \pdv[2]{V_{\col}}{\bar{\phi} (C)}{\phi (C^{\prime})} 
\right|_{\phi_0\{\phi^0_\alpha \}} 
A_{jC'}\Big|_{\phi_\alpha^0} \, ,
\hspace{10pt} i,j = 1, \dots , 2 N^2 \, .
\end{equation}
Here the ``bar'' symbols represent the complex conjugate. The mass matrix $\mathcal{M}_{\alpha\beta}$ is a matrix of dimension 
\[
    2 N^2 \times 2 N^2, 
\]
obtained from the multiplication of
\[
    [2 N^2 \times N_{\text{Loops}}] \times 
    [N_{\text{Loops}} \times N_{\text{Loops}}] \times 
    [N_{\text{Loops}} \times 2 N^2]
\]
matrices. It is essential that one sums over all $N_{\text{Loops}}$ in the equation above.\footnote{
    Indeed, it was shown in  \cite{Jevicki:1983hb} that if the sum is restricted to $N_\Omega$ loops, every non-zero eigenvalue of (\ref{LoopSpectrum}) is also an eigenvalue of (\ref{MasterSpectrumTrunc}), with $N_{\text{Loops}}$ replaced by $N_{\Omega}$. Except in the strong coupling phase of unitary matrix systems, the spectrum obtained simply on the basis of (\ref{LoopSpectrum}), with $\Omega^0$ a $(N_{\Omega} \times N_{\Omega} )$ matrix is not accurate.
    }
One obtains $N_{\Omega}$ \emph{nonzero} eigenvalues and $2 N^2 - N_{\Omega}$ (numerically) zero eigenvalues.

We now observe that 
\begin{equation*}
{\widehat\Omega}^0 (C,C') \equiv 
    \sum_{i}\bar{A}_{iC}\Big|_{\phi_\alpha^0}A_{iC'}\Big|_{\phi_\alpha^0} \, ,
    \qquad C,C' = 1, ... ,  N_{\rm Loops} \, ,
\end{equation*}
is a ($N_{\text{Loops}} \times N_{\text{Loops}}$) matrix which, in loop space, would include all loops up to length $4 l - 6$. In practice, it is not feasible to obtain such $\widehat{\Omega}$ directly in loop space, given the size of the truncations considered in this article (e.g., for $l=9\, ,\,\,  4 l - 6 = 30$). However, it can be generated from the master variables. 

The nonzero eigenvalues of (\ref{MasterSpectrumTrunc}) can then be matched with those of the loop space spectrum matrix
\begin{equation}\label{MatrixSpectrum}
     \mathcal{M}_{C,C^{\prime\prime}} =  \sum_{C^{\prime}=1}^{N_{\rm Loops} }\widehat{\Omega}_{0}(C, C^{\prime})V^{(2)} (C^{\prime}, C^{\prime\prime}) \, , \hspace{8pt} C,C^{\prime\prime} = 1, ... ,  N_{\rm Loops} \, . 
\end{equation}
This is a $N_{\text{Loops}} \times N_{\text{Loops}}$ matrix, and is expressed explicitly in terms of loop variables. It has $N_{\Omega}$ nonzero eigenvalues and $N_{\text{Loops}}  - N_{\Omega}$ (numerically) zero eigenvalues. Throughout, we have checked that the nonzero eigenvalues of (\ref{MasterSpectrumTrunc}) and (\ref{MatrixSpectrum}) are identical. 

\subsection{Optimization Procedure} 

A \texttt{Python} code was developed to obtain the computational results. The first step is to generate all the distinct single trace loops of a given length $l$. There are different ways of generating them, including using \texttt{Mathematica} and Polya theory. A simple procedure is to generate the list of $C_{l}^{n}$ combinations for $0 \le n \le l $, which index the position of (say) the matrix $M_2$ in the string (word) of $M_1$ and $M_2$ matrices, and then remove loops which are identical up to cyclic permutations. They are then indexed and stored as a list of arrays, eg. $ [1, 1, 1],  [1, 1, 2],  [1, 2, 2], [2, 2, 2] $, etc., and stacked for different lengths, to obtain the list of $N_{\Omega}$ and $N_{\operatorname{Loops}}$ loops described in \Cref{subsec:loop_trucation}, reproducing Table \ref{table:1}. The zeroth indexed element of the list is the empty array corresponding to $ \phi(0) = \Tr(\mathrm{I}) / N = 1$. It is fixed throughout.

The next step is to generate the loop joining matrix $\Omega$ 
\begin{eqnarray*}
 \Omega(c,c') = \sum_{a=1}^2 \pdv{\bar{\phi}(c)}{(M_a)_{ij}} \pdv{\phi(c')}{(M_a)_{ji}} = 
\sum_{c^{\prime\prime}=0}^{N_{\operatorname{Loops}}} j(c,c';c^{\prime \prime}) \phi(c^{\prime \prime})   , \hspace{8pt}        c,c' = 1, \dots, N_{\Omega} \, .
 \end{eqnarray*}
Here we use little `$c$' instead of capital `$C$' to emphasize the loop truncation: the loop joining matrix $\Omega$ now is a finite matrix of dimension $N_{\Omega} \times N_{\Omega}$ instead of an infinite dimensional matrix. The code implements explicitly the first equality in the above equation, recalling that\footnote{in the equation below `$\cdots$' stands for terms generated when the derivative does not act on the $M_1$ shown on the left hand side of the first line.}
\begin{eqnarray}\label{MatDeriv}
\frac{\partial}{\partial (M_1)_{ij}}  \Tr(\cdots  M_1 \cdots) &=&  \frac{\partial}{\partial (M_1)_{ij}}  
\Tr ( M_1\, g(\cdots M_1\cdots M_2\cdots ))+\cdots\cr 
&=& g_{ji} (\cdots M_1\cdots M_2\cdots) +\cdots
\end{eqnarray}
($g_{ij}$ is obtained by extracting $M_1$ from the loop). The joined loop $ \phi(c^{\prime\prime}) $ is identified, and the \emph{nonzero} joining coefficients $j(c, c'; c'')$  are stored. It should be emphasized that, through a joining process higher loops are generated, and the $N_\Omega \times N_\Omega$ matrix $\Omega$ contains higher loops up to length $N_{\rm{Loops}}$. It is this full set of loops contained in and generated by $\Omega$ that will participate in the optimization process.

A similar procedure is followed to generate $\omega$ defined through loop splitting
\begin{equation*}
\omega(c) = \sum_{a=1}^2 \frac{\partial}{\partial (M_a)_{ij} }  \frac{\partial \phi(c)}{\partial (M_a)_{ji}          } 
= \sum_{c',c''=0}^{N_{\operatorname{Loops}}} p(c;c',c'') \phi(c') \phi(c''), \hspace{8pt}    c = 1, ..., N_{\Omega}\, .
\end{equation*}
The split loops $ \phi(c'), \phi(c'') $ are identified, and the \emph{nonzero} splitting coefficients $p(c;c',c'')$  are stored. Since a given loop always splits into two smaller loops, $\omega$ only depends on the subset of $N_{\Omega}$ loop variables.

The master variables are the $N \times N$ matrices $M_1$ and $M_2$. For the minimization, due to the single trace nature of the invariant loops, $M_1$ is chosen diagonal and $M_2$ is an arbitrary $N  \times N$ hermitian matrix which we parametrize in the Lie algebra of $U(N)$
\begin{equation*}
 (M_1)_{ij}= \sum_{a=1}^N \mathfrak{a}_{aa} t^{aa}_{ij} \, , \qquad 
   (M_2)_{ij}= \sum_{a=1}^N \mathfrak{b}_{aa} t^{aa}_{ij} + \sum_{a < b}^N \mathfrak{b}_{ab} t^{ab}_{ij} +  \sum_{a > b}^N \mathfrak{b}_{ab} t^{ab}_{ij} \, .
\end{equation*}
Here $t^{ab}_{ij}$ $(a<b)$ is the set of real off-diagonal generators ($\sigma_1$ in the entries $(ij)$ and $(ji)$). $ t^{ab}_{ij}$ $(a>b)$ are the purely imaginary generators ($\sigma_2$ in the entries $(ij)$ and $(ji)$), and $t^{aa}_{ij} $ are the entries of a diagonal matrix.

In order to extract the explicit dependence on the powers of $N$ from the loops, they are defined as
\begin{equation}\label{NormLoops}
\phi(c) = \Tr ( \cdots M_1 \cdots  M_2  \cdots  M_1 \cdots  M_2  \cdots) /{N^{\frac{\rm{len}(c)}{2}+1}} \, ,
\end{equation}
where $\operatorname{len}(c)$ is length of the word $c$, i.e. the number of matrices in the loop.   

The function to be minimized is \eqref{MPPot} for matrix integrals or \eqref{Ham_S_2} for matrix quantum mechanics. The argument of the function is the real concatenated array $\mathfrak{a}_{aa}, a=1,...,N$ with the flattened matrix array $\mathfrak{b}_{ab}\, , a,b = 1,...,N$. At each iteration, for a given configuration of master variables, the $N_{\rm{Loops}}$ loops are evaluated from \eqref{NormLoops} and $\Omega$ and $\omega$ are evaluated with the values of the loops together with the coefficients $j(c, c'; c'')$ and $p(c; c',c'')$. Inversion of $\Omega$ is avoided by solving the relevant linear equations.

We used a standard \texttt{minimize} function from the \texttt{scipy.optimize} library. We used two methods, the \texttt{BFGS} and the \texttt{CG} methods. We found that the \texttt{BFGS} method is slightly faster, but for large loop truncations, the \texttt{CG} method is more stable. Both these methods require the evaluation of the gradient. This is achieved by calculating for each iteration the derivatives of the loops with respect to the master variables 
\begin{equation*}
\frac{\partial \phi(c)}{\partial \phi_\alpha} , \qquad 
\phi_\alpha \equiv (\mathfrak{a}_{aa},  \mathfrak{b}_{ab}) , \quad a,b=1,\dots,N\, .
\end{equation*}

The initial master variables configuration consists of a randomly generated real vector and of a randomly generated real matrix. For Fokker-Planck (and the underlying $c=0$) type systems, we have set as convergence criteria that the norm of the gradient vector becomes less than $\sqrt{N(N+1)} \,10^{-16}$. In other words, at convergence, a typical gradient vector element has norm of order $10^{-16}$. Convergence of the algorithm is remarkably stable, with the energy monotonically decreasing to zero in successive iterations. Depending on the size of the truncation, the energy at convergence is $\sim 10^{-24} -  10^{-31}$. The norm of the gradient components typically range from $\sim 10^{-15} -  10^{-20}$. At convergence, the Schwinger-Dyson equations 
are satisfied to typical accuracy $\sim 10^{-10} -  10^{-18}$. With a 3.0 GHz Mac, the codes take from about a few seconds for $N_{\text{Loops}} = 37 $, about two hours for $N_{\text{Loops}} = 2615 $ and more than a day for $N_{\text{Loops}} = 8923 $.

For spectra of the MQM ($c = 1$) systems discussed in \ref{sec:matrix-quantum-mechanics-and-spectrum}, we use $N = 94$ to generate large $N$ background. The initial master variables configuration again consists of a randomly generated real vector and of a randomly generated real matrix. The convergence criteria is that the norm of the gradient vector becomes less than $\sqrt{N(N+1)} \,10^{-16}$. Again, for $N_{\text{Loops}} = 37 $, converge is achieved in hours, while for $N_{\text{Loops}} = 8923$ convergence takes days.

For a given loop truncation size $N_{\text{Loops}}$, starting with the lowest $N$ satisfying $N(N+1) \ge N_{\text{Loops}}$ and increasing $N$, loop values are seen not to change much. Loop values quickly converge to their exact values (when known) as $N_{\text{Loops}}$ is increased. This is particularly so for the $N_{\Omega}$ small loops. This is evidenced, for example, in the matrix integrals case, in Table \ref{NewTable:2}.

\section{SD Models}
\label{sec:SD-models}

As a first application of the methods outlined above, we will study matrix integrals. For these models (in the decoupled case) there are exact analytic calculations which can be used to validate our numerical results. In all cases we are able to confirm that our numerical results are essentially exact for loops of lower length, with the accuracy falling off as we approach $L_{\rm max}$ which is the maximal length of loops admitted in the minimization. We note that all these models have a Hamiltonian quantum mechanical interpretation (Fokker-Planck). We also consider a variety of problems with two matrices to demonstrate the methods. It will be clear that the extensions to more matrices proceed in the same way, without difficulty.

Numerically we are evaluating the integral \eqref{loopvev} with the action $S$ given by
\begin{equation}
S = V(M_1) + V(M_2) + k \Tr (M_1M_2) \, ,
\end{equation}
where the potential is given by
\begin{equation}
V(M)=\frac{1}{2} \Tr (M^2) + \frac{g_3}{\sqrt{N}}{\rm Tr} M^3 + \frac{g_4}{N}{\rm Tr} M^4 \, .
\end{equation}
When the coupling $k = 0$ we will refer to the system as a ``single matrix system'' and when $k \ne 0$ as a ``two-matrix system''. Note however, that for both types of systems we evaluate mixed loops $\phi (C)$ obtained by tracing products involving both matrices $M_1$ and $M_2$, so that even when $k = 0$ the problem is still a multi matrix problem.


\subsection{Single Matrix Systems}

\paragraph{Free theory.} 
The potential of the zero dimensional model is
\begin{eqnarray}
V(M) = \frac{1}{2}{\rm Tr} M^2 \, ,
\end{eqnarray}
so that we have set $g_3=g_4=0$. We also set $k = 0$. The effective potential is
\begin{eqnarray}
V_{\rm eff} = \frac{1}{8}\omega\Omega^{-1}\omega + \frac{1}{8}{\rm Tr} (M_1^2) + \frac{1}{8}{\rm Tr} (M_2^2)
- \frac{1}{2} N^2 \, .
\end{eqnarray}

\paragraph{Quartic theory.} 
The potential of the corresponding zero dimensional model is
\begin{eqnarray}
V(M) = \frac{1}{2}{\rm Tr} M^2+ \frac{g_4}{N}{\rm Tr} M^4 \, ,
\end{eqnarray}
so that we have set $k = g_3 = 0$. The effective potential is
\begin{eqnarray}
V_{\rm eff} & = & \frac{1}{8} \omega\Omega^{-1}\omega
 + \left( \frac{1}{8} - 2 g_4 \right) {\rm Tr} (M_1^2)
 + \left( \frac{1}{8} - 2 g_4 \right) {\rm Tr} (M_2^2)
 - \frac{1}{2} N^2 \cr\cr
& & - \frac{g_4}{N} \left( {\rm Tr} (M_1)^2 + {\rm Tr}(M_2)^2 \right)
 + \frac{g_4}{N} \left({\rm Tr} (M_1^4) + {\rm Tr} (M_2^4) \right) \cr\cr
& & + \frac{2g_4^2}{N^2} \left({\rm Tr} (M_1^6)+{\rm Tr} (M_2^6)\right) \, .
\end{eqnarray}

\paragraph{Cubic model.} 
We use
\begin{eqnarray}
V(M) = \frac{1}{2} {\rm Tr} M^2 + \frac{g_3}{\sqrt{N}}{\rm Tr} M^3 \, ,
\end{eqnarray}
so that we have set $k = g_4 = 0$. The effective potential is
\begin{eqnarray}
V_{\rm eff} & = & \frac{1}{8} \omega\Omega^{-1}\omega
 + \frac{1}{8} {\rm Tr} (M_1^2) + \frac{1}{8} {\rm Tr} (M_2^2)
 + \frac{3 g_3}{4\sqrt{N}}{\rm Tr} (M_1^3) + \frac{3g_3}{4\sqrt{N}}{\rm Tr} (M_2^3)
 - \frac{1}{2} N^2 \cr\cr
& & + \frac{9g_3^2}{8N}{\rm Tr} (M_1^4) + \frac{9g_3^2}{8N}{\rm Tr} (M_2^4)
 - \frac{3g_3}{2\sqrt{N}} {\rm Tr} (M_1) - \frac{3g_3}{2\sqrt{N}}{\rm Tr} (M_2) \, .
\end{eqnarray}

\subsection{Two-Matrix Systems}

\paragraph{Quadratic model.} 
The potential of the zero dimensional model is
\begin{eqnarray}
V(M) = \frac{1}{2}{\rm Tr} M^2 \, ,
\end{eqnarray}
so that we have set $g_3 = g_4 = 0$. In this case we keep $ k \ne 0$. The effective potential is
\begin{eqnarray}
V_{\rm eff} = \frac{1}{8} \omega\Omega^{-1}\omega 
 + \frac{k^2+1}{8} {\rm Tr} (M_1^2) 
 + \frac{k^2+1}{8} {\rm Tr} (M_2^2)
 + \frac{k}{4}{\rm Tr} (M_1 M_2) 
 - \frac{1}{2} N^2 \, .
\end{eqnarray}

\paragraph{Coupled cubic two-matrix model.} 
We keep $k > 0$ and we set
\begin{eqnarray}
V(M) = \frac{1}{2}{\rm Tr} M^2 + \frac{g_3}{\sqrt{N}}{\rm Tr} M^3 \, ,
\end{eqnarray}
so that we have set $g_4 = 0$. The effective potential is
\begin{eqnarray}
V_{\rm eff} 
& = & \frac{1}{8} \omega\Omega^{-1}\omega 
    + \frac{1+k^2}{8} {\rm Tr} (M_1^2) 
    + \frac{1+k^2}{8}{\rm Tr} (M_2^2)
    + \frac{k}{4}{\rm Tr} (M_1 M_2) 
    - \frac{1}{2} N^2 
    \cr\cr
& & + \frac{3 g_3}{4\sqrt{N}} {\rm Tr} (M_1^3)
    + \frac{3 g_3}{4\sqrt{N}} {\rm Tr} (M_2^3)
    + \frac{9 g_3^2}{8N} {\rm Tr} (M_1^4)
    + \frac{9 g_3^2}{8N} {\rm Tr} (M_2^4)
    \cr\cr
& & + \frac{3 g_3 k}{4\sqrt{N}}{\rm Tr} (M_1^2 M_2)
    + \frac{3 g_3 k}{4\sqrt{N}}{\rm Tr} (M_1 M_2^2)
    - \frac{3 g_3}{2\sqrt{N}}{\rm Tr} (M_1)
    - \frac{3 g_3}{2\sqrt{N}}{\rm Tr} (M_2) \,.
\end{eqnarray}

\subsection{Results}

In tables 2-10 we present results for SD models, obtained with $l=9$ and $N=94$.\footnote{
    In the following tables the traces are scaled by appropriate factors according to \eqref{NormLoops} so that the results of loops are independent of the matrix size $N$.
    } 
The results show that for small loops we essentially obtain the exact answer. For larger loops (with more than $l$ matrices in the trace, in the notation of Section \ref{sec:methods}) the results are less accurate, but even for the longest loops accuracy is typically always better than 5\%.

\begin{table}[htb!]
\small
    \begin{center}
    \begin{tabular}{||c|l|l||l|l||l|l||} 
    \hline
    $g_4$           & \multicolumn{2}{c||}{$0$}  & \multicolumn{2}{c||}{$1$}   &  \multicolumn{2}{c||}{$10$} \\ [0.5ex] 
    \hline
      & Exact       & Numerical & Exact       & Numerical & Exact       & Numerical \\ [0.5ex] 
    \hline\hline
    ${\rm Tr}(M_1^2)$    &1  &0.9999565 &0.3125 &0.3139285 &0.113752 &0.1166236\\ 
    \hline
    ${\rm Tr}(M_1 M_2)$ &0  &0.0002630 &0         &$8.846\times 10^{-6}$ &0 &0.0002037\\ 
    \hline
    ${\rm Tr}(M_1^4)$    &2  &1.9998894 &0.171875 &0.1732885 &0.0221562 &0.0230570\\ 
    \hline
    ${\rm Tr}(M_1^2 M_2^2)$ &1  &0.9999058 &0.0976563 &0.09810712 &0.0129395 &0.0132793\\ 
    \hline
    ${\rm Tr}(M_1^6)$ &5  &4.9997429 &0.113281 &0.11462865 &0.00513368 &0.0054013\\ 
    \hline
    ${\rm Tr}(M_1^2 M_2^4)$ &2  &1.9998074 &0.0537109 &0.05394868 &0.00252031 &0.0025842\\ 
    \hline
    ${\rm Tr}(M_1^4 M_2^2)$ &2  &1.9997449 &0.0537109 &0.05415388 &0.00252031 &0.0026246\\ 
    \hline
    ${\rm Tr}(M_1^4 M_2^4)$ &4  &3.9994546 &0.029541 &0.02977492 &0.000490897 &0.0005107\\ 
    \hline
    \end{tabular}
    \caption{Results for the $g_3 = k = 0$ as $g_4$ is varied. The table shows loops with lengths $\le 8$.}
    \end{center}
    \end{table}
    
\begin{table}[htb!]
\small
    \begin{center}
    \begin{tabular}{||c||l|l||l|l||l|l||} 
    \hline
    $g_4$           & \multicolumn{2}{c||}{$0$}  & \multicolumn{2}{c||}{$1$}   &  \multicolumn{2}{c||}{$10$} \\ [0.5ex] 
    \hline
    & Exact           & Numerical     & Exact           & Numerical    & Exact            & Numerical \\ [0.5ex] 
    \hline\hline
    ${\rm Tr}(M_1^{10})$       &42  &41.898479 &0.0629883 &0.064156 &0.000350126 &0.000376\\ 
    \hline
    ${\rm Tr}(M_1^8 M_2^2)$ &14  &13.986907 &0.0256348 &0.026020 &0.000148213 &0.000157\\ 
    \hline
    ${\rm Tr}(M_1^5M_2^5)$  &0  &0.014260 &0.0 &1.030$\times 10^{-5}$ &0.0 &-5.322$\times 10^{-7}$\\ 
    \hline
    ${\rm Tr}(M_1^{12})$       &132 &130.63810 &0.050354 &0.051447 &0.0000978653 &0.000106\\ 
    \hline
    ${\rm Tr}(M_1^6M_2^6)$  &25  &25.130274 &0.0128326 &0.012996 &2.635$\times 10^{-5}$ &2.779$\times 10^{-5}$\\ 
    \hline
    ${\rm Tr}(M_1^{14})$       &429  &417.95735 &0.041458 &0.042414 &2.816$\times 10^{-5}$ &3.053$\times 10^{-5}$\\ 
    \hline
    ${\rm Tr}(M_1^4 M_2^{10})$ &84  &84.278562 &0.0108261 &0.010975 &7.757$\times 10^{-6}$ &8.111$\times 10^{-6}$\\ 
    \hline
    ${\rm Tr}(M_1^{16})$            &1430  &1359.7222 &0.0349121 &0.035647 &8.283$\times 10^{-6}$ &8.941$\times 10^{-6}$\\ 
    \hline
    \end{tabular}
    \caption{Results for the $g_3 = k =0$ as $g_4$ is varied. The loops shown have lengths $\ge 10$ and $\le 16$.}
    \end{center}
    \end{table}
    
\begin{table}[htb!]
\small
    \begin{center}
    \begin{tabular}{||c||l|l||l|l||l|l||} 
    \hline
    $g_3$           & \multicolumn{2}{c||}{$0.01$}  & \multicolumn{2}{c||}{$0.025$}   &  \multicolumn{2}{c||}{$0.05$} \\ [0.5ex]
    \hline 
    & Exact       & Numerical & Exact       & Numerical & Exact       & Numerical \\ [0.5ex] 
    \hline\hline
    ${\rm Tr}(M_1^2)$            &1.00363         &1.00363 &1.02358              &1.02358 &1.11155 &1.11155\\ 
    \hline
    ${\rm Tr}(M_1 M_2)$         &0.000906539  &0.000909 &0.00589         &0.00589 &0.027799 &0.027798\\ 
    \hline
    ${\rm Tr}(M_1^4)$            &2.02182         &2.02182 &2.14427              &2.14429 &2.73465 &2.73462\\ 
    \hline
    ${\rm Tr}(M_1^2 M_2^2)$ &1.00727         &1.00727 &1.04771             &1.04771 &1.23554 &1.23554\\ 
    \hline
    ${\rm Tr}(M_1^6)$            &5.10951         &5.10950 &5.73859             &5.73869 &9.1135 &9.1131\\ 
    \hline
    ${\rm Tr}(M_1^2 M_2^4)$ &2.02915         &2.02916 &2.19482             &2.19485 &3.0397 &3.03967\\ 
    \hline
    ${\rm Tr}(M_1^4 M_2^2)$ &2.02915         &2.02915 &2.19482             &2.19485 &3.0397 &3.03967\\ 
    \hline
    ${\rm Tr}(M_1^4 M_2^4)$ &4.08777         &4.08779 &4.59788             &4.59810 &7.47831 &7.47913\\ 
    \hline
    \end{tabular}
    \caption{Results for $g_4 = k = 0$ as $g_3$ is varied. The table shows loops with lengths $\le 8$.}
    \end{center}
    \end{table}
    
\begin{table}[htb!]
\small
    \begin{center}
    \begin{tabular}{||c||l|l||l|l||l|l||} 
    \hline
    $g_3$           & \multicolumn{2}{c||}{$0.01$}  & \multicolumn{2}{c||}{$0.025$}   &  \multicolumn{2}{c||}{$0.05$} \\ [0.5ex]
    \hline 
     & Exact           & Numerical     & Exact           & Numerical    & Exact            & Numerical \\ [0.5ex]  
    \hline\hline
    ${\rm Tr}(M_1^{10})$       &44.3207 &44.2172 &58.4568 &58.3197 &156.937 &156.449\\ 
    \hline
    ${\rm Tr}(M_1^8 M_2^2)$ &14.5658  &14.5441 &17.9569 &17.9432 &39.9365 &39.8981\\ 
    \hline
    ${\rm Tr}(M_1^5M_2^5)$  &0.207449  &0.2311 &1.4793 &1.5072 &10.4007 &10.4477\\ 
    \hline
    ${\rm Tr}(M_1^{12})$       &142.267 &140.851 &207.031 &204.9967 &732.18 &723.9593\\ 
    \hline
    ${\rm Tr}(M_1^6M_2^6)$  &26.1071 &26.2563 &32.9314 &33.1235 &83.0558 &83.4798\\ 
    \hline
    ${\rm Tr}(M_1^{14})$       &473.759 &462.0533 &767.231 &748.6339 &3570.97 &3476.2509\\ 
    \hline
    ${\rm Tr}(M_1^4 M_2^{10})$ &89.6086 &89.8709 &125.347 &125.4809 &429.168 &427.3324\\ 
    \hline
    ${\rm Tr}(M_1^{16})$            &1623.11 &1546.8709 &2943.81 &2808.6594 &17971.5 &17086.8 \\ 
    \hline
    \end{tabular}
    \caption{Results for $g_4 = k = 0$ as $g_3$ is varied. The loops shown have lengths $\ge 10$ and $\le 16$.}
    \end{center}
    \end{table}

\begin{table}[htb!]
\small
    \begin{center}
    \begin{tabular}{||c||l|l||l|l||l|l||} 
    \hline
    $k$           & \multicolumn{2}{c||}{$0$}  & \multicolumn{2}{c||}{$\frac{1}{2}$}   &  \multicolumn{2}{c||}{$\frac{3}{4}$} \\ [0.5ex]
    \hline
        & Exact       & Numerical & Exact       & Numerical & Exact       & Numerical \\ [0.5ex]  
    \hline\hline
    ${\rm Tr}(M_1^2)$            &1  &0.9999565 &1.33333     &1.333362  &2.28571  &2.28559\\ 
    \hline
    ${\rm Tr}(M_1 M_2)$         &0  &0.0002630 &-0.666667  &-0.666619 &-1.71429 &-1.71416\\ 
    \hline
    ${\rm Tr}(M_1^4)$            &2  &1.999885   &3.55556     &3.555304  &10.449    &10.447\\ 
    \hline
    ${\rm Tr}(M_1^2 M_2^2)$ &1  &0.9999058 &2.22222     &2.221766  &8.16327 &8.16228\\ 
    \hline
    ${\rm Tr}(M_1^6)$            &5  &4.9997429 &11.8519     &11.8497    &59.7085 &59.6988\\ 
    \hline
    ${\rm Tr}(M_1^2 M_2^4)$ &2  &1.9998074 &6.51852     &6.51612    &44.035   &44.0271\\ 
    \hline
    ${\rm Tr}(M_1^4 M_2^2)$ &2 &1.9997449  &6.51852     &6.51603    &44.035   &44.0271\\ 
    \hline
    ${\rm Tr}(M_1^4 M_2^4)$ &4  &3.9994546 &19.9506     &19.9388    &256.0     &255.9\\ 
    \hline
    \end{tabular}
    \caption{Results for $g_3 =g_4 = 0$ as $k$ is varied. The table shows loops with length $\le 8$.}
    \end{center}
    \end{table}
    
\begin{table}[htb!]
\small
    \begin{center}
    \begin{tabular}{||c||l|l||l|l||l|l||} 
    \hline
    $k$           & \multicolumn{2}{c||}{$0$}  & \multicolumn{2}{c||}{$\frac{1}{2}$}   &  \multicolumn{2}{c||}{$\frac{3}{4}$} \\ [0.5ex]
    \hline
                  & Exact           & Numerical     & Exact           & Numerical    & Exact            & Numerical \\ [0.5ex] 
    \hline\hline
    ${\rm Tr}(M_1^{10})$       &42 &41.898479 &176.988 &178.1885 &2620.35 &2649.95\\ 
    \hline
    ${\rm Tr}(M_1^8 M_2^2)$ &14  &13.986907 &88.4938 &88.7473 &1856.08 &1869.81\\ 
    \hline
    ${\rm Tr}(M_1^5M_2^5)$  &0  &0.014260 &-61.2346 &-61.2468 &-1605.73 &-1610.73\\ 
    \hline
    ${\rm Tr}(M_1^{12})$       &132 &130.63810 &741.663 &759.5443 &18823.7 &19572.3\\ 
    \hline
    ${\rm Tr}(M_1^6M_2^6)$  &25 &25.130274 &263.111 &264.3137 &11215.9 &11339.8\\ 
    \hline
    ${\rm Tr}(M_1^{14})$       &429 &417.95735 &3213.87 &3378.6847 &139833 &151758\\ 
    \hline
    ${\rm Tr}(M_1^4 M_2^{10})$ &84 &84.278562 &1170.09 &1169.2822 &84618.9 &85304.5\\ 
    \hline
    ${\rm Tr}(M_1^{16})$            &1430 &1359.7222 &14283.9 &15516.1101 &1.0654$\times 10^6$ &1.2204$\times 10^6$ \\ 
    \hline
    \end{tabular}
    \caption{Results for $g_3 = g_4 = 0$ as $k$ is varied. The loops shown have lengths $\ge 10$ and $\le 16$.}
    \end{center}
    \end{table}


\begin{table}[htb!]
    \begin{center}
    \begin{tabular}{||c|c|c|c|c|c|c||} 
    \hline
    $N_{\text{Loops}}$ & $93$     & $93$     & $261$   & $261$   & $801$     & $801$    \\ [0.5ex] \hline
    $N_{\Omega}$       & $23$     & $23$     & $37$    & $37$    & $57$      & $57$     \\ [0.5ex] \hline
    $N$                & $10$     & $12$     & $16$    & $18$    & $28$      & $30$     \\ [0.5ex] 
    \hline\hline
    ${\rm Tr}(M_1^2)$    & 1.398558& 	 1.398550   & 1.398602 	& 1.398606  & 1.398608 	& 1.398608  \\ 
    \hline
    ${\rm Tr}(M_1 M_2)$ & 0.724017 &	 0.724014  & 0.724058 &	 0.724062 &  0.724064 &	 0.724064  \\ 
    \hline
    ${\rm Tr}(M_1^4)$    &  4.078131 	& 4.075882  &  4.083250 &	 4.083690 &  4.083917 	& 4.083920   \\ 
    \hline
    ${\rm Tr}(M_1^2 M_2^2)$ &  2.584743 &	 2.584708   &  2.588234 &	 2.588669 &  2.588752 	& 2.588754   \\ 
    \hline
    ${\rm Tr}(M_1^6)$ & 14.627466 	& 13.991341  & 15.388399 	& 15.430672    & 15.477357 &	 15.477725   \\ 
    \hline
    ${\rm Tr}(M_1^2 M_2^4)$ &  8.479602 	& 8.622227   &  8.692989 	 &8.740663  & 8.739481 	& 8.739705   \\ 
    \hline
    ${\rm Tr}(M_1^4 M_2^2)$ & 8.404034 	& 8.284243  & 8.688555 	& 8.727894   &  8.738072 & 	 8.738460  \\ 
    \hline
    ${\rm Tr}(M_1^4 M_2^4)$ & 56.718119 &	 52.025168    &  63.460122 	& 64.980069  & 66.088900 	& 66.155167   \\ 
    \hline
    \end{tabular}
    \caption{Results for $g_3=0.05, \, k=0.5 $ for $l=5,6,7$ and increasing $N$. Note that the length of the last loop is $>7$.}
    \label{NewTable:2}
    \end{center}
    \end{table}    

\begin{table}[htb!]
\small
    \begin{center}
    \begin{tabular}{||c||c|c|c|c|c|c||} 
    \hline
    $N_{\text{Loops}}$  & $37$     & $93$  & $261$    & $801$ & $2615$    & $8923 $\\[0.5ex] 
    \hline
    $N_{\Omega}$         & $15$  & $23$    & $37$  & $57$   & $93$  & $153$ \\ [0.5ex] 
    \hline\hline
    $\varepsilon_1$    &0.49236511 &0.49236370 &0.49236470   &0.49236443&  0.49236443  &0.49236463\\ 
    \hline
    $\varepsilon_2$ &0.49342716 &0.49343864  &   0.49344212      &0.49344752&0.49344586 & 0.49344757    \\ 
    \hline
    \end{tabular}
    \caption{The two lowest lying states for $k = g_4 = 0$ and $g_3=0.05/3=0.0167$.}
    \label{SpecTable:1}
    \end{center}
    \end{table}

\begin{table}[htb!]
\small
    \begin{center}
    \begin{tabular}{||c||c|c|c|c|c|c||} 
    \hline
    $N_{\text{Loops}}$  & $37$  & $93$  & $261$  & $801$ & $2615$  & $8923$ \\ [0.5ex] 
    \hline
    $N_{\Omega}$        & $15$  & $23$  & $37 $  & $57 $ & $93  $  & $153 $ \\ [0.5ex] 
    \hline\hline
    $\varepsilon_1$     &5.44592002 &5.25941037 & 5.13024606 &5.20001164 & 5.18903527  & 5.15228747 \\ 
    \hline
    $\varepsilon_2$     &5.81626188 &5.33316170 & 5.41489587 &5.40011675 & 5.36702168  & 5.33021481 \\ 
    \hline
    \end{tabular}
    \caption{The two lowest lying states for $k = g_3 = 0$ and $g_4=10$.}
    \label{SpecTable:2}
    \end{center}
    \end{table}

\clearpage
\newpage

\section{Matrix Quantum Mechanics\ Spectrum}
\label{sec:matrix-quantum-mechanics-and-spectrum}

In this section, we consider multi-matrix quantum mechanical models, which includes the evaluation of (Wilson) loop expectation values at large $N$, ground state energies at large $N$, and the spectrum of fluctuations (which corresponds to $N^0$, i.e. order $1$). The coupled  two-matrix Hamiltonian reads
\begin{eqnarray} \label{eq:2M-Hamiltonian}
H=\frac{1}{2}{\rm Tr}(\Pi_1^2+\Pi_2^2)+{V(M_1,M_2)} \, ,
\end{eqnarray} 
with a potential \begin{eqnarray} \label{eq:2M-interaction}
V(M_1, M_2)=\frac{1}{2}\Tr(M_1^2+M_2^2)
+k\Tr(M_1 M_2)+ \frac{g_4}{N}\Tr(M_1^4+M_2^4) \, .
\end{eqnarray} 
We note, that the SD models featured in the previous section are also of this form, but an effective potential containing double trace couplings. Consequently the numerical methods for evaluating the stationary points and the spectrum apply to both with no difference in degree of difficulty. Likewise, our methods can be applied to multi-matrix models with arbitrary number of matrices. We describe evaluation of the spectrum for the above case. For normalization purposes, we will also give numerical (and analytical) one-matrix results.

\subsection{Fluctuations}

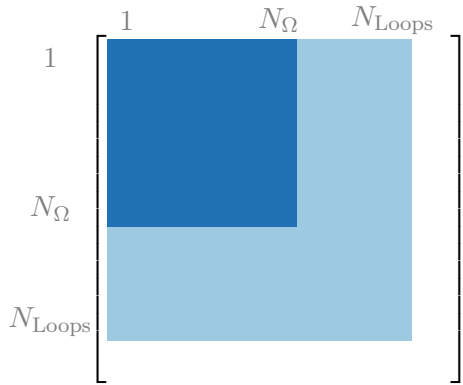
\begin{figure}[t!]
    \begin{center}
        
     \scalebox{1}{\begin{tikzpicture}[
    every left delimiter/.style={xshift=1ex},
    every right delimiter/.style={xshift=-1ex}
]
\draw[gray] (-2,2.5) node{1};
\draw[gray] (0,2.5) node{$N_{\Omega}$};
\draw[gray] (1.5,2.5) node{$N_{\rm Loops}$};
\draw[gray] (-3,2) node{1};
\draw[gray] (-3,0) node{$N_{\Omega}$};
\draw[gray] (-3,-1.5) node{$N_{\rm Loops}$};

\matrix[
    nodes={minimum size=5mm},
    matrix of math nodes,
    left delimiter={[},
    right delimiter={]}]
  {
    \node[fill=myblue1] {}; & \node[fill=myblue1] {}; & \node[fill=myblue1] {}; & \node[fill=myblue1] {};  & \node[fill=myblue1] {}; & \node[fill=myblue2] {}; & \node[fill=myblue2] {}; & \node[fill=myblue2] {}; & {}\\
    \node[fill=myblue1] {}; & \node[fill=myblue1] {}; & \node[fill=myblue1] {}; & \node[fill=myblue1] {};  & \node[fill=myblue1] {}; & \node[fill=myblue2] {}; & \node[fill=myblue2] {}; & \node[fill=myblue2] {}; & {}\\
    \node[fill=myblue1] {}; & \node[fill=myblue1] {}; & \node[fill=myblue1] {}; & \node[fill=myblue1] {};  & \node[fill=myblue1] {}; & \node[fill=myblue2] {}; & \node[fill=myblue2] {}; & \node[fill=myblue2] {}; & {}\\
    \node[fill=myblue1] {}; & \node[fill=myblue1] {}; & \node[fill=myblue1] {}; & \node[fill=myblue1] {};  & \node[fill=myblue1] {}; & \node[fill=myblue2] {}; & \node[fill=myblue2] {}; & \node[fill=myblue2] {}; & {}\\
    \node[fill=myblue1] {}; & \node[fill=myblue1] {}; & \node[fill=myblue1] {}; & \node[fill=myblue1] {};  & \node[fill=myblue1] {}; & \node[fill=myblue2] {}; & \node[fill=myblue2] {}; & \node[fill=myblue2] {}; & {}\\
    \node[fill=myblue2] {}; & \node[fill=myblue2] {}; & \node[fill=myblue2] {}; & \node[fill=myblue2] {};  & \node[fill=myblue2] {}; & \node[fill=myblue2] {}; & \node[fill=myblue2] {}; & \node[fill=myblue2] {}; & {}\\
    \node[fill=myblue2] {}; & \node[fill=myblue2] {}; & \node[fill=myblue2] {}; & \node[fill=myblue2] {};  & \node[fill=myblue2] {}; & \node[fill=myblue2] {}; & \node[fill=myblue2] {}; & \node[fill=myblue2] {}; & {}\\
    \node[fill=myblue2] {}; & \node[fill=myblue2] {}; & \node[fill=myblue2] {}; & \node[fill=myblue2] {};  & \node[fill=myblue2] {}; & \node[fill=myblue2] {}; & \node[fill=myblue2] {}; & \node[fill=myblue2] {}; & {}\\
    {} & {} & {} & {} & {} \\
  };
  
\end{tikzpicture}}  

    \end{center}
    \caption{Illustration of $\Omega$ truncation for spectrum calculation.}
    \label{fig:Omega_illustration}
\end{figure}

The general strategy for the spectrum calculation of small fluctuations in loop space, which holds for all multi-matrix quantum mechanics in the limit of large $N$, has been described in previous sections, and is easily implementable on a computer. In this subsection we explain the subtly about truncation further. As in the optimization procedure, we truncate the infinite dimensional loop space. Let $l$ denote the chosen loop length truncation, and $N_{\Omega}$ denote the number of loops whose lengths are less or equal to~$l$. $\Omega$ then is a $N_{\Omega}$ by $N_{\Omega}$ matrix which involves $N_{\rm Loops}$ loops in total, and the loop of maximal length it contains is $L = 2 l -2 $. Sending $\Omega \rightarrow N^{-2} \Omega$, we obtain the collective Hamiltonian
\begin{equation} \label{eq:H_col}
H_{\col}=\frac{1}{2 N^2} \sum_{c=1}^{N_{\Omega}} P^{\dagger}(c) \Omega(c, c^{\prime}) P(c^{\prime}) + N^2 V_{\rm col}[\phi] \, ,
\end{equation}
where
\begin{equation} \label{eq:V_col}
    V_{\rm col}[\phi] = \frac{1}{8} \sum_{c=1}^{N_{\Omega}} \bar{\omega}(c) \Omega^{-1}(c, c^{\prime}) \omega(c^{\prime}) + v[\phi] \, .
\end{equation}
Here $N^2 v[\phi]$ is the original potential written in terms of loops. We then expand $H_{\col}$ to order $\mathcal{O}(1)$, which is accomplished by shifting loop variables around the ground state
\begin{equation}
    \phi(C) = \phi_{0}(C) + \frac{1}{N} \eta (C), \quad P(C) = N p(C), \quad \text{where } C = 1, \dots , N_{\text{Loops}} \, ,
\end{equation}
and omitting the constant background terms.  We use Capital $C$ to emphasize that here the loop labels run from $1$ to $N_{\rm Loops}$, instead of $N_{\Omega}$, because $V_{\col}$ in equation \eqref{eq:V_col} involves $N_{\rm Loops}$ independent loops. Accordingly there are also $N_{\rm Loops}$ canonical conjugates, which requires us to use the $N_{\rm Loops}$ by $N_{\rm Loops}$ dimensional loop joining matrix denoted as $\widehat{\Omega}$. The choice of the loop joining matrix is illustrated in Figure \ref{fig:Omega_illustration}, where the deep blue blocks are used for truncation in $V_{\col}$, and the deep plus light blue blocks, i.e. $\widehat{\Omega}$, are used in the following $H_{\col}^{(2)}$. 

The Taylor expansion then gives
\begin{equation} \label{eq:H_coll_quadratic}
    H_{\col}^{(2)} = \sum_{C=1}^{N_{\rm Loops}} 
    \left(\frac{1}{2} p^{\dagger} (C) \widehat{\Omega}_{0}(C, C^{\prime}) p(C^\prime) 
    +  \frac{1}{2} \bar{\eta}(C) V^{(2)}_{0}(C, C^{\prime}) \eta(C^{\prime})\right) \, ,
\end{equation}
in which $V^{(2)}_0$ is the Hessian matrix of $V_{\col}$ at the ground state. We note that all elements in $\Omega$ are linear functions of loops. Their second derivatives therefore are 0, and hence do not contribute to $V^{(2)}_0$. For the same reason, the second derivative of $v[\phi]$ vanishes if it only contains single trace terms. 

To evaluate the spectrum, one can diagonalize the kinetic term in \eqref{eq:H_coll_quadratic} and then solve the eigenvalues of the resulting mass matrix. As pointed out earlier, we see that using $\widehat{\Omega}$ also resolves the mismatch of the dimensions between $\Omega_0$ and $V^{(2)}_0$. Since $\widehat{\Omega}_0$ is positive definite, one can perform a canonical transformation
\begin{equation}
    \eta \rightarrow \sqrt{\widehat{\Omega}_0} \: \eta, \quad p \rightarrow \sqrt{\widehat{\Omega}_0^{-1}} \: p \, .
\end{equation}
The spectrum is then given, in terms of the \emph{nonzero} eigenvalues of the spectrum matrix $\widehat{\Omega}_0 V^{(2)}_0$, by
\begin{equation} \label{eq:spectrum_expression}
    \varepsilon_n = \left[\operatorname{eig}_{n}
        \left(\sum_{C^{\prime}=1}^{N_{\rm Loops}}\widehat{\Omega}_{0}(C, C^{\prime})V^{(2)}_0 (C^{\prime}, C^{\prime\prime})\right)\right]^{1/2}, \quad n \in \mathbb{Z}^{+} \, ,
\end{equation}
where $\operatorname{eig}_n$ denotes the $n$th nonzero eigenvalue. Here $n$ starts from $1$ instead of $0$, because the zero mode is excluded in the definition of $\Omega$ (or $\widehat{\Omega}$). The spectrum is independent of truncation loop length $l$, as we will see in the following concrete examples. In principle, the size of the spectrum one can obtain is equal to the $N_{\Omega}$. Besides, there are precisely $N_{\rm Loops} - N_{\Omega}$ zero eigenvalues of $\widehat{\Omega}_0 V^{(2)}_0$, therefore the truncation scheme automatically projects out the higher modes. As $l$ is increased, higher modes are included, and one is able to obtain higher frequencies.

Depending on the size of the truncation and particularly for multi-matrix systems, it is not feasible in general to obtain $\widehat{\Omega}_0 $ directly in loop space. But $\widehat{\Omega}_0 $ can be always be constructed with master variables. The spectrum equation  \eqref{eq:spectrum_expression} then takes the form
\begin{equation}
    \varepsilon_{n} = \left[\operatorname{eig}_{n}
        \left(
            \sum_{C^\prime=1}^{N_{\rm Loops}}
            \sum_{a}\sum_{i,j}
            \pdv{\bar{\phi}(C)}{(\bar{M}_{a})_{ij}} 
            \pdv{\phi(C^{\prime})}{(M_{a})_{ji}} 
            V^{(2)}_0(C^\prime, C^{\prime \prime})
        \right)\right]^{1/2} \, .
\end{equation}
For example, the spectrum that is presented in Section 5.3 is obtained in the background of 801 (this is the number of loops for a cut off
of $L_{\rm max} =2\times 7-2$) loops, whose value is determined by the master field after minimization.
This results in a spectrum eigenvalue matrix of size $\approx 10^3 \times 10^3$.
Note also that 401 947 loops are effectively included in the computation of $\widehat{\Omega}_0$.
This is only possible through the use of master variables and a direct loop space evaluation would not be possible.
It is visible that the spectrum calculation in terms of master fields should give the same results, except for different numbers of zero eigenvalues of $\widehat{\Omega}_0 V^{(2)}_{0}$.

We observe that in the free theory cases, interestingly, one can actually obtain exact results by working with a smaller $V^{(2)}_0$ matrix, whose matrix indices range only from $1$ to $N_{\Omega}$, and correspondingly with the smaller $\Omega_0$ matrix. This has been verified in both one- and two-matrix quantum mechanics. This is no longer the case the moment coupling constants are switched on.

\subsection{One-Matrix Example}

\begin{figure}[t!]
\begin{center}
    \begin{subfigure}[a]{0.48\linewidth}
        
     \scalebox{0.85}{\begin{tikzpicture}
\begin{axis}[
    cycle list={
        {cycle1, smooth, mark=*, mark size=1.5pt, line width=0.8pt},
        {cycle2, smooth, mark=triangle*, mark size=2pt, line width=0.8pt},
        {cycle3, smooth, mark=square*, mark size=1.5pt, line width=0.8pt},
        {cycle4, smooth, mark=diamond*, mark size=2pt, line width=0.8pt},
        {cycle5, smooth, mark=pentagon*, mark size=2pt, line width=0.8pt},
        {cycle6, smooth, mark=halfcircle*, mark size=1.5pt, line width=0.8pt}
    },
    xmin=1, xmax=20,
    ymin=0, 
    xlabel={$n$},
    ylabel={$\varepsilon_n$},
    xtick={1,5,10,15,20},
    grid=minor,
    legend style={
        at={(0.25,0.95)}, 
        anchor=north,
    },
]

\addplot table[x=level,y=-0.05] {data_1M_spectrum_first_30_levels_numerical.dat};
\addlegendentry{$g_4=-0.05$}

\addplot table[x=level,y=0] {data_1M_spectrum_first_30_levels_numerical.dat};
\addlegendentry{$g_4=0$}

\addplot table[x=level,y=1] {data_1M_spectrum_first_30_levels_numerical.dat};
\addlegendentry{$g_4=1$}

\addplot table[x=level,y=5] {data_1M_spectrum_first_30_levels_numerical.dat};
\addlegendentry{$g_4=5$}

\addplot table[x=level,y=10] {data_1M_spectrum_first_30_levels_numerical.dat};
\addlegendentry{$g_4=10$}

\end{axis}
\end{tikzpicture}}  

        \caption{Frequencies versus level numbers with different couplings.}
        \label{fig:1M-spectrum-vs-levels}
    \end{subfigure}
    \hfill
    \begin{subfigure}[a]{0.48\linewidth}
        
     \scalebox{0.85}{    
\begin{tikzpicture}

\begin{axis}[
    cycle list/Paired-12,
    xmin=-1, xmax=10,
    ymin=0,
    xlabel={$g_4$},
    ylabel={$\varepsilon_1$},
    xtick={0,2,4,6,8,10},
    grid=minor,
    extra x ticks={-0.0750264},
    extra x tick style={grid=major},
    extra x tick labels={},
    legend style={
        at={(0.8,0.2)}, 
        anchor=north,
    },
]

\addplot+[
    smooth,
    mark=,
    line width=1.0pt,
] table {data_1M_spectrum_1st_frequency_analytical.dat};
\addlegendentry{analytic}

\addplot+[
    only marks,
    mark=*,
    mark size=1.5pt,
] table {data_1M_spectrum_1st_frequency_numerical.dat};
\addlegendentry{numeric}

\end{axis}

\end{tikzpicture}}  

        \caption{The first frequency $\varepsilon_1$ versus the quartic coupling $g_4$.}
        \label{fig:1M-spectrum-vs-g4}
    \end{subfigure}
    \caption{Spectrum of the large $N$ Hermitian one-matrix quantum mechanics.}
\end{center}
\end{figure}
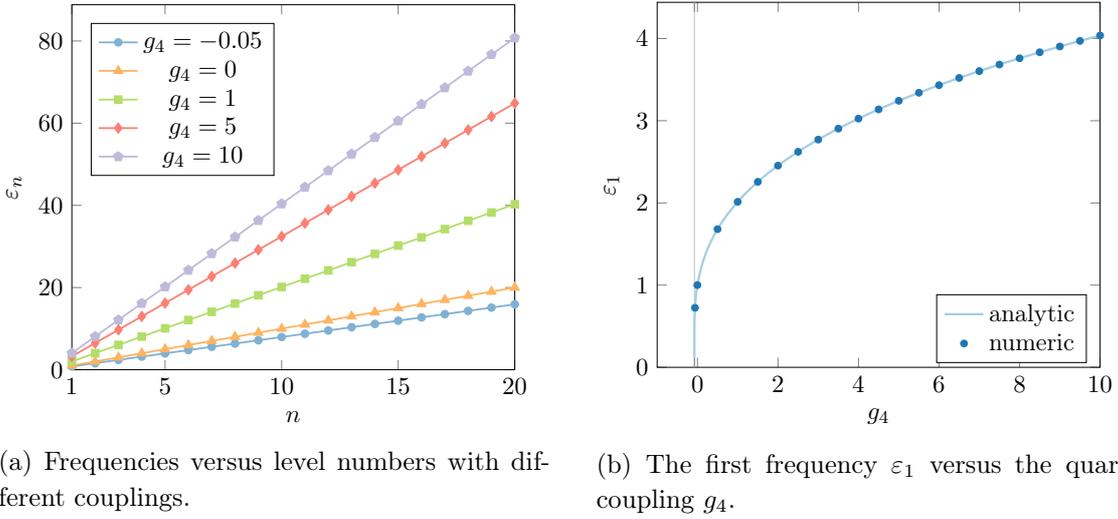

We then proceed to employ our general strategy to compute the spectrum of the hermitian one-matrix quantum mechanics: 
\begin{equation}
    H = - \frac{1}{2}\Tr(\pdv[2]{}{M}) + \frac{1}{2} \Tr(M^2) + \frac{g_4}{N} \Tr(M^4) \, .
\end{equation}
In this simple case the all loop variables are real, and are labeled by a nonnegative integer $n$ so that $\phi(n) = \Tr(M^n)/N^{n/2+1}$. The loop joining and splitting have components
\begin{equation}
    \Omega(n, m) = N^{-2} \, n \, m \, \phi(n+m-2),
    \qquad
    \omega(n) = n \sum_{m=0}^{n-2} \phi(m) \phi(n - m - 2) \, .
\end{equation}
Some analytical results including the spectrum formula are summarized in Appendix \ref{appendix:one-matrix-analytical}. With the analytical loop values \eqref{eq:loop-analytical} we can also obtain the spectrum using a computer. A \texttt{Mathematica} program was developed for the spectrum calculation. The one-matrix case has the simple feature that $N_{\Omega} = l$ and also $N_{\rm Loops} = L_{\rm max}$, which provides us a canonical example to illustrate our general strategy. To calculate the spectrum using the general strategy at loop length truncation $l=6$, for example, there are totally $L_{\rm max} = 10$ loops contained in $V_{\col}$. $\Omega_0$ then is a $6 \times 6$ matrix, but $\widehat{\Omega}_0$ and $V^{(2)}_{0}$ are both $10 \times 10$ matrices. In Table \ref{tab:1M-spectrum} we present both the exact \eqref{eq:1M_spectrum_analytical} and numerical low lying spectrum results for various $g_4$, showing excellent agreement. These results are also relevant to the following two-matrix example. In Figure \ref{fig:1M-spectrum-vs-levels} we present several numerical results of spectrum $\varepsilon_n$ versus level indices $n$. They all fit into a straight line, revealing the simple relation $\varepsilon_n = n \varepsilon_1$, which is also a property predicted by the exact result \eqref{eq:1M_spectrum_analytical}. When $g_4=0$, the model reduces to decoupled harmonic oscillators, therefore we have $\varepsilon_n = n$, as is verified. For other couplings we present the first level frequency $\varepsilon_{1}$ in Figure \ref{fig:1M-spectrum-vs-g4}. The exact and numerical results again agree very well, including the critical region $g \sim g_{\operatorname{c}} = - 1 / 3\sqrt{2}\pi$.

\begin{table}[t!]
\begin{center}
    
\pgfplotstableread{data_1M_spectrum_exact.dat}\loadedtable
\pgfplotstabletypeset[
    every head row/.style={before row={\hline \multicolumn{1}{||l}{} \\ \hline}, after row=\hline \hline},
    every last row/.style={after row=\hline},
    columns={g4},
    columns/g4/.style={column name={$g_4$}, column type=||ll, string type},
    skip rows between index={6}{10},
]\loadedtable
\pgfplotstabletypeset[
    every head row/.style={before row={\hline \multicolumn{5}{||c||}{exact results} \\ \hline}, after row=\hline \hline},
    every last row/.style={after row=\hline},
    columns={0,0.1,1,10,100},
    display columns/0/.style={precision=3, column type=||l},
    display columns/1/.style={precision=3, column type=|l},
    display columns/2/.style={precision=3, column type=|l},
    display columns/3/.style={precision=3, column type=|l},
    display columns/4/.style={precision=3, column type=|l||},
    skip rows between index={6}{10}
]\loadedtable
\begin{filecontents*}{data_1M_spectrum_numerical.dat}
0	0period1	1	10	100
1.	1.2233983392477634	2.013716587614586	4.037014753674751	8.55726738145476
2.	2.446796672438611	4.027433171267471	8.074029441060562	17.11453453742662
3.	3.6701950213897816	6.0411542454434795	12.111096317504485	25.671965663471703
4.	4.893593361394828	8.054874923747484	16.148158188681286	34.229380690861376
5.	6.116994071705055	10.069122771717042	20.189029523596762	42.79752904938965
6.	7.34039398198363	12.083184227718041	24.228508422416574	51.361716250114895
7.	8.564498523944174	14.125005858622718	28.391296345134023	60.236767423400195
8.	9.788175247909756	16.1492703742182	32.47435249035024	68.90964009685842
9.	11.074452780593758	18.65803677741313	37.97196166856931	80.83389385202246
10.	12.30907827599196	20.761375747258136	42.2767134095506	90.01077482252475
\end{filecontents*}
\pgfplotstabletypeset[
    every head row/.style={before row={\hline \multicolumn{5}{c||}{numerical results} \\ \hline}, after row=\hline \hline},
    every last row/.style={after row=\hline},
    columns={0,0period1,1,10,100},
    display columns/0/.style={precision=3, column type=l},
    display columns/1/.style={column name=0.1, precision=3, column type=|l},
    display columns/2/.style={precision=3, column type=|l},
    display columns/3/.style={precision=3, column type=|l},
    display columns/4/.style={precision=3, column type=|l||},
    skip rows between index={6}{10}
]{data_1M_spectrum_numerical.dat}

    \caption{One-matrix spectrum.}
    \label{tab:1M-spectrum}
\end{center}
\end{table}

\subsection{Two-Matrix example}

By applying the same strategy, and using the numerical minimization results, we are also able to evaluate the spectrum of the hermitian two-matrix quantum mechanics \eqref{eq:2M-Hamiltonian} and \eqref{eq:2M-interaction}, using a \texttt{Mathematica} program. 

At first we present in Table \ref{tab:E_gs_0_comparison} the comparison of the ground state energies, in which the two-matrix results are obtained from numerical minimization, and the one-matrix results are obtained from the analytical expression \eqref{eq:1M-Egs_0}. The two-matrix model in $c=0$ cases doubles the one-matrix model interaction, and one expects that the ground state energy should therefore be twice that of the corresponding one-matrix result. Table \ref{tab:E_gs_0_comparison} exhibits excellent agreement, verifying the expectation.
\begin{table}[t!]
\begin{center}
    
\pgfplotstableread{data_E_gs_0_comparison.dat}\loadedtable
\pgfplotstabletypeset[
    every head row/.style={before row={\hline}, after row=\hline \hline},
    every last row/.style={after row=\hline},
    columns={g4, 2x1Mexact, 2M(lmax9data)},
    columns/g4/.style={column name=$g_4$, column type=||l|},
    columns/2x1Mexact/.style={column name={$2\times$one-matrix}, column type=l|, precision=5},
    columns/2M(lmax9data)/.style={column name={two-matrix}, column type=l||, precision=5}
]\loadedtable

    \caption{$E_{\operatorname{gs}}^{(0)}$ comparison.}
    \label{tab:E_gs_0_comparison}
\end{center}
\end{table}

As for the spectrum computation, the multi-matrix models lose the simple feature $N_{\Omega}=l$, and a rapid growth of $N_{\Omega}$ has already been observed in the two-matrix case. Based on the $l=9$ ($l=7$) numerical minimization data for interacting (free) case, we present some spectrum results evaluated at truncation $l=4, 5, 6, 7$ in \Cref{tab:2M-spectrum_lmax_4}, \ref{tab:2M-spectrum_lmax_5}, \ref{tab:2M-spectrum_lmax_6}, \ref{tab:2M-spectrum_lmax_7} and \ref{tab:2M-spectrum_lmax_7_part2}. Taking the $l=6$ case for example, one has $N_{\Omega}=37$, so that $\Omega$ in $V_{\operatorname{eff}}$ is $37 \times 37$. On the other hand, the fact $L = 10$ requires us to use the $N_{\text{Loops}}=261$ dimensional $\widehat{\Omega}_0$ and $V^{(2)}_{0}$, which implies a large amount ($261-37=224$) of the zero eigenvalues of $\widehat{\Omega}_0 V^{(2)}_{0}$. This was verified by our \texttt{Mathematica} program. Comparing the two-matrix spectrum results, we see that as $l$ is increased, low lying spectrum are stable and convergent, and higher level frequencies are obtained.

\begin{figure}[htb!]
\begin{center}
    \begin{tabular}{rl}
\begin{tikzpicture}
\begin{axis}[
    title={$k=0$ and $g_4=0$},
    xmin=1, xmax=37,
    ymin=0, 
    xlabel={$n$},
    ylabel={$\varepsilon_n$},
    xtick={2,5,9,15,23,37},
    ytick={1,2,3,4,5,6},
    xmajorgrids,
    width={\dimexpr0.5\textwidth-7pt},
]
\addplot[cycle1, only marks, mark=*, mark size=1.5pt] table[x=n,y=run1] {data_2M_spectrum_lmax_6_for_plot.dat};
\end{axis}
\end{tikzpicture}
&
\begin{tikzpicture}
    \begin{axis}[
        title={$k=0$ and $g_4=0.1$},
        xmin=1, xmax=37,
        ymin=0, 
        xlabel={$n$},
        ylabel={$\varepsilon_n$},
        xtick={2,5,9,15,23,37},
        ytick={2,4,6,8,10,12,14},
        yticklabel pos=upper,
        xmajorgrids,
        width={\dimexpr0.5\textwidth-7pt},
    ]
    \addplot[cycle2, only marks, mark=*, mark size=1.5pt] table[x=n,y=run2] {data_2M_spectrum_lmax_6_for_plot.dat};
    \end{axis}
    \end{tikzpicture}
\\
\begin{tikzpicture}
    \begin{axis}[
        title={$k=0.5$ and $g_4=0$},
        xmin=1, xmax=37,
        ymin=0, 
        xlabel={$n$},
        ylabel={$\varepsilon_n$},
        xtick={2,5,9,15,23,37},
        ytick={1,2,3,4,5,6,7,8},
        xmajorgrids,
        width={\dimexpr0.5\textwidth-7pt},
    ]
    \addplot[cycle3, only marks, mark=*, mark size=1.5pt] table[x=n,y=run6] {data_2M_spectrum_lmax_6_for_plot.dat};
    \end{axis}
    \end{tikzpicture}
&
\begin{tikzpicture}
    \begin{axis}[
        title={$k=0.5$ and $g_4=1$},
        xmin=1, xmax=37,
        ymin=0, 
        xlabel={$n$},
        ylabel={$\varepsilon_n$},
        xtick={2,5,9,15,23,37},
        ytick={2,4,6,8,10,12,14},
        yticklabel pos=upper,
        xmajorgrids,
        width={\dimexpr0.5\textwidth-7pt},
    ]
    \addplot[cycle4, only marks, mark=*, mark size=1.5pt] table[x=n,y=run8] {data_2M_spectrum_lmax_6_for_plot.dat};
    \end{axis}
    \end{tikzpicture}
\\
\end{tabular}
    \caption{Two-matrix spectrum plots and degeneracy lifting due to $g_4$ and $k$.}
    \label{fig:2M-spectrum}
\end{center}    
\end{figure}
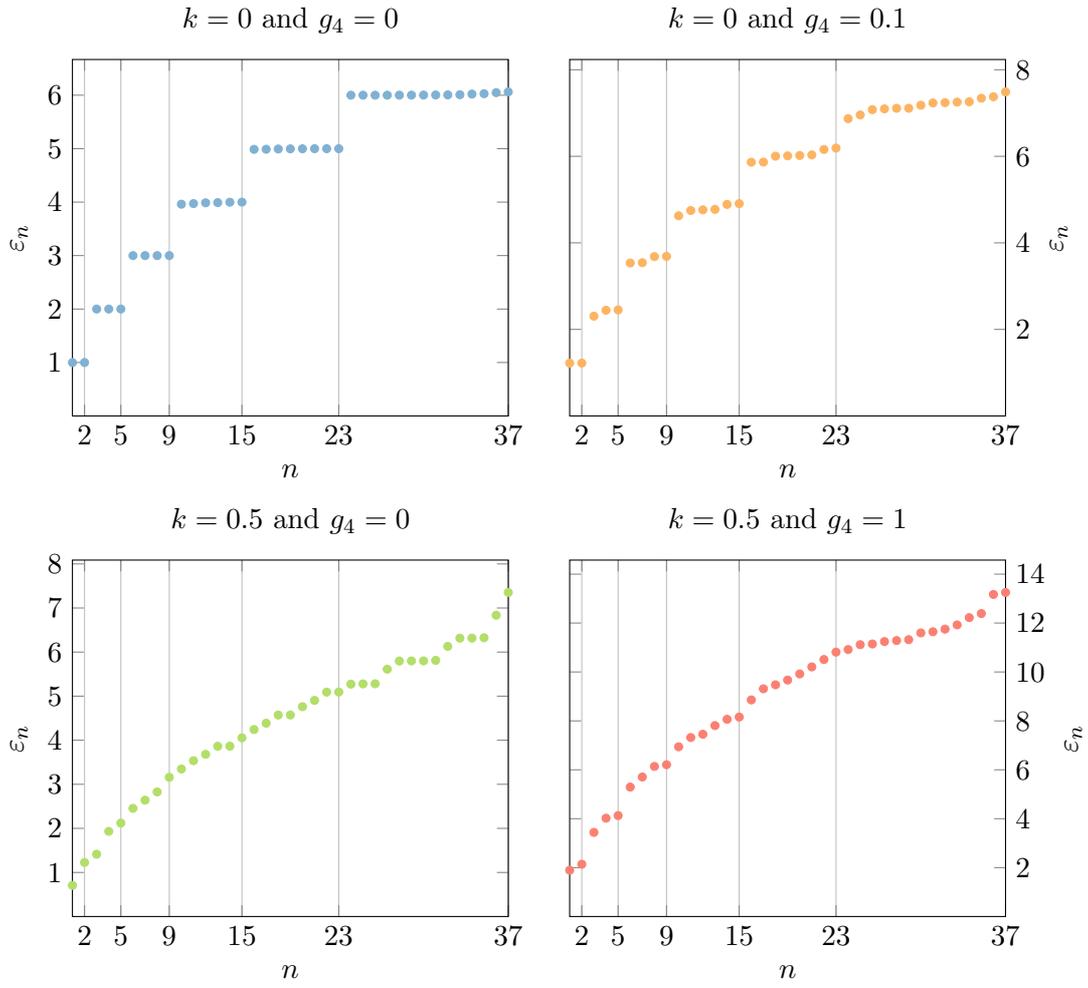

Let us now examine in more detail the primary characteristics of the two-matrix spectra. As the results show, in free theory, $k = g_4 = 0$, the spectrum values coincide with the one-matrix case. Besides, a high degeneracy pattern is observed. Let $\mathcal{N}(n)$ denote the number of inequivalent loops of length $n$. The degeneracy at level $n$ is precisely equal to $\mathcal{N}(n)$. For instance, at level 2 the degeneracy is 3, since there are 3 independent loops, namely $\Tr(M_1 M_1)$, $\Tr(M_1 M_2)$, and $\Tr(M_2 M_2)$. As we turn on the quartic coupling $g_4$ while fixing $k = 0$, the degeneracy is lifted slightly, and a different degeneracy pattern is obtained. Comparing with the one-matrix results Table \ref{tab:1M-spectrum}, one can obviously see that the two-matrix spectrum contains the corresponding one-matrix spectrum as a subset, and with a degeneracy number 2. These subsets are filled in light blue colors in each $k = 0$ and $g \neq 0$ column of the two-matrix spectrum results. When turning on coupling $k$, the degeneracy patterns are nearly destroyed. The lifting of the degeneracy due to $k$ and $g_4$ is presented in Figure \ref{fig:2M-spectrum} based on data in Table \ref{tab:2M-spectrum_lmax_6}.  We conjecture that these properties are \emph{universal} for multi-matrix models.

\begin{table}[htb!]
\begin{center}
    
\pgfplotstableread{data_2M_spectrum_lmax_4.dat}\loadedtable
\pgfplotstabletypeset[
    every head row/.style={output empty row, after row=\hline},
    every last row/.style={after row=\hline},
    every row no 2/.style={before row=\hline \hline},
    every row no 4/.style={before row=\hline},
    every row no 7/.style={before row=\hline},
    every row no 11/.style={before row=\hline},
    every row no 17/.style={before row=\hline},
    every row no 25/.style={before row=\hline},
    columns={run},
    columns/run/.style={column type=||ll, string type},
]\loadedtable
\pgfplotstabletypeset[
    every head row/.style={output empty row, after row=\hline},
    every last row/.style={after row=\hline},
    every row no 2/.style={before row=\hline \hline},
    every row no 4/.style={before row=\hline},
    every row no 7/.style={before row=\hline},
    every row no 11/.style={before row=\hline},
    every row no 17/.style={before row=\hline},
    every row no 25/.style={before row=\hline},
    columns={1},
    columns/1/.style={precision=3, column type=||c}
]\loadedtable
\pgfplotstabletypeset[
    every head row/.style={output empty row, after row=\hline},
    every last row/.style={after row=\hline},
    every row no 1/.style={after row={\rowcolor{myblue2}}},
    every row no 2/.style={before row=\hline \hline, after row={\rowcolor{myblue2}}},
    every row no 4/.style={before row=\hline, after row={\rowcolor{myblue2}}},
    every row no 5/.style={after row={\rowcolor{myblue2}}},
    every row no 7/.style={before row=\hline},
    every row no 8/.style={after row={\rowcolor{myblue2}}},
    every row no 9/.style={after row={\rowcolor{myblue2}}},
    every row no 11/.style={before row=\hline},
    every row no 14/.style={after row={\rowcolor{myblue2}}},
    every row no 15/.style={after row={\rowcolor{myblue2}}},
    every row no 17/.style={before row=\hline},
    every row no 22/.style={after row={\rowcolor{myblue2}}},
    every row no 23/.style={after row={\rowcolor{myblue2}}},
    every row no 25/.style={before row=\hline},
    every row no 36/.style={after row={\rowcolor{myblue2}}},
    every row no 37/.style={after row={\rowcolor{myblue2}}},
    columns={2,3,4,5},
    columns/2/.style={precision=3, column type =||l|},
    columns/3/.style={precision=3, column type=l|},
    columns/4/.style={precision=3, column type=l|},
    columns/5/.style={precision=3, column type=l||}
]\loadedtable
\pgfplotstabletypeset[
    every head row/.style={output empty row, after row=\hline},
    every last row/.style={after row=\hline},
    every row no 2/.style={before row=\hline \hline},
    every row no 4/.style={before row=\hline},
    every row no 7/.style={before row=\hline},
    every row no 11/.style={before row=\hline},
    every row no 17/.style={before row=\hline},
    every row no 25/.style={before row=\hline},
    columns={6,7,8},
    columns/6/.style={precision=3, column type=l},
    columns/7/.style={precision=3, column type=|l},
    columns/8/.style={precision=3, column type=|l||}
]\loadedtable

    \caption{Two-matrix spectrum with loop length truncation $l=4$.}
    \label{tab:2M-spectrum_lmax_4}
\end{center}
\end{table}

\begin{table}[htb!]
\begin{center}
    
\pgfplotstableread{data_2M_spectrum_lmax_5.dat}\loadedtable
\pgfplotstabletypeset[
    every head row/.style={output empty row, after row=\hline},
    every last row/.style={after row=\hline},
    every row no 2/.style={before row=\hline \hline},
    every row no 4/.style={before row=\hline},
    every row no 7/.style={before row=\hline},
    every row no 11/.style={before row=\hline},
    every row no 17/.style={before row=\hline},
    every row no 25/.style={before row=\hline},
    columns={run},
    columns/run/.style={column type=||ll, string type},
]\loadedtable
\pgfplotstabletypeset[
    every head row/.style={output empty row, after row=\hline},
    every last row/.style={after row=\hline},
    every row no 2/.style={before row=\hline \hline},
    every row no 4/.style={before row=\hline},
    every row no 7/.style={before row=\hline},
    every row no 11/.style={before row=\hline},
    every row no 17/.style={before row=\hline},
    every row no 25/.style={before row=\hline},
    columns={1},
    columns/1/.style={precision=3, column type=||c}
]\loadedtable
\pgfplotstabletypeset[
    every head row/.style={output empty row, after row=\hline},
    every last row/.style={after row=\hline},
    every row no 1/.style={after row={\rowcolor{myblue2}}},
    every row no 2/.style={before row=\hline \hline, after row={\rowcolor{myblue2}}},
    every row no 4/.style={before row=\hline, after row={\rowcolor{myblue2}}},
    every row no 5/.style={after row={\rowcolor{myblue2}}},
    every row no 7/.style={before row=\hline},
    every row no 8/.style={after row={\rowcolor{myblue2}}},
    every row no 9/.style={after row={\rowcolor{myblue2}}},
    every row no 11/.style={before row=\hline},
    every row no 14/.style={after row={\rowcolor{myblue2}}},
    every row no 15/.style={after row={\rowcolor{myblue2}}},
    every row no 17/.style={before row=\hline},
    every row no 22/.style={after row={\rowcolor{myblue2}}},
    every row no 23/.style={after row={\rowcolor{myblue2}}},
    every row no 25/.style={before row=\hline},
    every row no 36/.style={after row={\rowcolor{myblue2}}},
    every row no 37/.style={after row={\rowcolor{myblue2}}},
    columns={2,3,4,5},
    columns/2/.style={precision=3, column type =||l|},
    columns/3/.style={precision=3, column type=l|},
    columns/4/.style={precision=3, column type=l|},
    columns/5/.style={precision=3, column type=l||}
]\loadedtable
\pgfplotstabletypeset[
    every head row/.style={output empty row, after row=\hline},
    every last row/.style={after row=\hline},
    every row no 2/.style={before row=\hline \hline},
    every row no 4/.style={before row=\hline},
    every row no 7/.style={before row=\hline},
    every row no 11/.style={before row=\hline},
    every row no 17/.style={before row=\hline},
    every row no 25/.style={before row=\hline},
    columns={6,7,8},
    columns/6/.style={precision=3, column type=l},
    columns/7/.style={precision=3, column type=|l},
    columns/8/.style={precision=3, column type=|l||}
]\loadedtable

    \caption{Two-matrix spectrum with loop length truncation $l=5$.}
    \label{tab:2M-spectrum_lmax_5}
\end{center}
\end{table}

\begin{table}[htb!]
\begin{center}
    
\pgfplotstableread{data_2M_spectrum_lmax_6.dat}\loadedtable
\pgfplotstabletypeset[
    every head row/.style={output empty row, after row=\hline},
    every last row/.style={after row=\hline},
    every row no 2/.style={before row=\hline \hline},
    every row no 4/.style={before row=\hline},
    every row no 7/.style={before row=\hline},
    every row no 11/.style={before row=\hline},
    every row no 17/.style={before row=\hline},
    every row no 25/.style={before row=\hline},
    columns={run},
    columns/run/.style={column type=||ll, string type},
]\loadedtable
\pgfplotstabletypeset[
    every head row/.style={output empty row, after row=\hline},
    every last row/.style={after row=\hline},
    every row no 2/.style={before row=\hline \hline},
    every row no 4/.style={before row=\hline},
    every row no 7/.style={before row=\hline},
    every row no 11/.style={before row=\hline},
    every row no 17/.style={before row=\hline},
    every row no 25/.style={before row=\hline},
    columns={1},
    columns/1/.style={precision=3, column type=||c}
]\loadedtable
\pgfplotstabletypeset[
    every head row/.style={output empty row, after row=\hline},
    every last row/.style={after row=\hline},
    every row no 1/.style={after row={\rowcolor{myblue2}}},
    every row no 2/.style={before row=\hline \hline, after row={\rowcolor{myblue2}}},
    every row no 4/.style={before row=\hline, after row={\rowcolor{myblue2}}},
    every row no 5/.style={after row={\rowcolor{myblue2}}},
    every row no 7/.style={before row=\hline},
    every row no 8/.style={after row={\rowcolor{myblue2}}},
    every row no 9/.style={after row={\rowcolor{myblue2}}},
    every row no 11/.style={before row=\hline},
    every row no 14/.style={after row={\rowcolor{myblue2}}},
    every row no 15/.style={after row={\rowcolor{myblue2}}},
    every row no 17/.style={before row=\hline},
    every row no 22/.style={after row={\rowcolor{myblue2}}},
    every row no 23/.style={after row={\rowcolor{myblue2}}},
    every row no 25/.style={before row=\hline},
    every row no 36/.style={after row={\rowcolor{myblue2}}},
    every row no 37/.style={after row={\rowcolor{myblue2}}},
    columns={2,3,4,5},
    columns/2/.style={precision=3, column type =||l|},
    columns/3/.style={precision=3, column type=l|},
    columns/4/.style={precision=3, column type=l|},
    columns/5/.style={precision=3, column type=l||}
]\loadedtable
\pgfplotstabletypeset[
    every head row/.style={output empty row, after row=\hline},
    every last row/.style={after row=\hline},
    every row no 2/.style={before row=\hline \hline},
    every row no 4/.style={before row=\hline},
    every row no 7/.style={before row=\hline},
    every row no 11/.style={before row=\hline},
    every row no 17/.style={before row=\hline},
    every row no 25/.style={before row=\hline},
    columns={6,7,8},
    columns/6/.style={precision=3, column type=l},
    columns/7/.style={precision=3, column type=|l},
    columns/8/.style={precision=3, column type=|l||}
]\loadedtable

    \caption{Two-matrix spectrum with loop length truncation $l=6$.}
    \label{tab:2M-spectrum_lmax_6}
\end{center}
\end{table}

\begin{table}[htb!]
\small
\begin{center}
    
\pgfplotstableread{data_2M_spectrum_lmax_7.dat}\loadedtable
\pgfplotstabletypeset[
    every head row/.style={output empty row, after row=\hline},
    every last row/.style={after row=\hline},
    every row no 2/.style={before row=\hline \hline},
    every row no 4/.style={before row=\hline},
    every row no 7/.style={before row=\hline},
    every row no 11/.style={before row=\hline},
    every row no 17/.style={before row=\hline},
    every row no 25/.style={before row=\hline},
    every row no 39/.style={before row=\hline},
    columns={run},
    columns/run/.style={column type=||ll, string type},
    skip rows between index={39}{59}
]\loadedtable
\pgfplotstabletypeset[
    every head row/.style={output empty row, after row=\hline},
    every last row/.style={after row=\hline},
    every row no 2/.style={before row=\hline \hline},
    every row no 4/.style={before row=\hline},
    every row no 7/.style={before row=\hline},
    every row no 11/.style={before row=\hline},
    every row no 17/.style={before row=\hline},
    every row no 25/.style={before row=\hline},    
    every row no 39/.style={before row=\hline},
    columns={1},
    columns/1/.style={precision=3, column type=||c},
    skip rows between index={39}{59}
]\loadedtable
\pgfplotstabletypeset[
    every head row/.style={output empty row, after row=\hline},
    every last row/.style={after row=\hline},
    every row no 1/.style={after row={\rowcolor{myblue2}}},
    every row no 2/.style={before row=\hline \hline, after row={\rowcolor{myblue2}}},
    every row no 4/.style={before row=\hline, after row={\rowcolor{myblue2}}},
    every row no 5/.style={after row={\rowcolor{myblue2}}},
    every row no 7/.style={before row=\hline},
    every row no 8/.style={after row={\rowcolor{myblue2}}},
    every row no 9/.style={after row={\rowcolor{myblue2}}},
    every row no 11/.style={before row=\hline},
    every row no 14/.style={after row={\rowcolor{myblue2}}},
    every row no 15/.style={after row={\rowcolor{myblue2}}},
    every row no 17/.style={before row=\hline},
    every row no 22/.style={after row={\rowcolor{myblue2}}},
    every row no 23/.style={after row={\rowcolor{myblue2}}},
    every row no 25/.style={before row=\hline},
    every row no 36/.style={after row={\rowcolor{myblue2}}},
    every row no 37/.style={after row={\rowcolor{myblue2}}},
    every row no 39/.style={before row=\hline},
    every row no 41/.style={before row={\rowcolor{myblue2}}, after row={\rowcolor{myblue2}}},
    columns={2,3,4,5},
    columns/2/.style={precision=3, column type =||l|},
    columns/3/.style={precision=3, column type=l|},
    columns/4/.style={precision=3, column type=l|},
    columns/5/.style={precision=3, column type=l||},
    skip rows between index={39}{59}
]\loadedtable
\pgfplotstabletypeset[
    every head row/.style={output empty row, after row=\hline},
    every last row/.style={after row=\hline},
    every row no 2/.style={before row=\hline \hline},
    every row no 4/.style={before row=\hline},
    every row no 7/.style={before row=\hline},
    every row no 11/.style={before row=\hline},
    every row no 17/.style={before row=\hline},
    every row no 25/.style={before row=\hline},
    every row no 39/.style={before row=\hline},
    columns={6,7,8},
    columns/6/.style={precision=3, column type=l},
    columns/7/.style={precision=3, column type=|l},
    columns/8/.style={precision=3, column type=|l||},
    skip rows between index={39}{59}
]\loadedtable

    \caption{Two-matrix spectrum with loop length truncation $l=7$.}
    \label{tab:2M-spectrum_lmax_7}
\end{center}
\end{table}

\begin{table}[htb!]
\begin{center}
    
\pgfplotstableread{data_2M_spectrum_lmax_7.dat}\loadedtable
\pgfplotstabletypeset[
    every head row/.style={output empty row, after row=\hline},
    every last row/.style={after row=\hline},
    every row no 2/.style={before row=\hline \hline},
    columns={run},
    columns/run/.style={column type=||ll, string type},
    skip rows between index={2}{39}
]\loadedtable
\pgfplotstabletypeset[
    every head row/.style={output empty row, after row=\hline},
    every last row/.style={after row=\hline},
    every row no 2/.style={before row=\hline \hline},
    columns={1},
    columns/1/.style={precision=3, column type=||c},
    skip rows between index={2}{39}
]\loadedtable
\pgfplotstabletypeset[
    every head row/.style={output empty row, after row=\hline},
    every last row/.style={after row=\hline},
    every row no 2/.style={before row=\hline \hline},
    every row no 19/.style={after row={\rowcolor{myblue2}}},
    every row no 20/.style={after row={\rowcolor{myblue2}}},
    columns={2,3,4,5},
    columns/2/.style={precision=3, column type =||l|},
    columns/3/.style={precision=3, column type=l|},
    columns/4/.style={precision=3, column type=l|},
    columns/5/.style={precision=3, column type=l||},
    skip rows between index={2}{39}
]\loadedtable
\pgfplotstabletypeset[
    every head row/.style={output empty row, after row=\hline},
    every last row/.style={after row=\hline},
    every row no 2/.style={before row=\hline \hline},
    columns={6,7,8},
    columns/6/.style={precision=3, column type=l},
    columns/7/.style={precision=3, column type=|l},
    columns/8/.style={precision=3, column type=|l||},
    skip rows between index={2}{39}
]\loadedtable

    \caption{Two-matrix spectrum with loop length truncation $l=7$ continued.}
    \label{tab:2M-spectrum_lmax_7_part2}
\end{center}
\end{table}

\clearpage
\newpage

\section{Conclusions}
\label{sec:conclusions}

We have applied previously developed numerical, master field methods to solve a variety of coupled two-matrix models. These include matrix integrals and matrix quantum mechanics systems, with the fact that they all have a representation in terms of collective Hamiltonians with (Wilson) loops as dynamical variables. The collective loop space representation provides (nonlinear) Wilson loop equations, which in the former case (matrix integral) are equivalent to Schwinger-Dyson (Migdal-Makeenko) equations. Solution as it was understood in the original minimization schemes is to be accomplished subject to loop space (Schwarz) inequalities which define the constrained minimization procedure. The collective method includes not only the form of the large $N$ Hamiltonian and potential but also provides a complete set of inequalities, which we give in explicit determinate form). The associated (constrained) minimization of the large $N$ problem is implemented through a master field, as explained in detail in the text (and also in original works). The numerical solution of the nonlinear large $N$ stationary point not only gives the leading large $N$ Wilson loop background, it also establishes in concrete terms the existence of the master field. This existence. was questioned through the years in various works. Constrained minimization, which is accomplished with fairly large number of (loop) variables ($\sim 10^4$) is seen to give essentially exact results for large $N$ expectation values, ground state energy and low lying spectra. This size can be further increased for added precision. The formulation, and methods developed are such that there is no difference in having larger number of matrix variables. The large $N$ dynamics is evaluated in loop space, which is parametrized the same way (and easily computer augmented). Regarding further works, and  applications of the methods developed in this work, one can contemplate many physical problems requiring understanding through matrix models range from confinement \cite{Hanada:2019czd} to cosmology \cite{Brahma:2021tkh}. More specifically, extensions to supersymmetric versions of the multi-matrix QM can be considered. The four matrix BMN quantum mechanics (at large $N$) is clearly accessible by this scheme, and we plan to present results in future work. Considering the two matrix case, whose solution is accomplished in this work, an interesting application to entanglement \cite{Das:2020jhy}, thermodynamics and the matrix thermofield double (TFD) states. In general this requires a study of QM defined on the Schwinger-Keldysh contour \cite{Schwinger:1960qe,Keldysh:1964ud}. The TFD state and the corresponding wave functional of the $O(N)$ vector theory based on the collective field formulation was recently studied in \cite{Jevicki:2021ddf}. Adjusting our optimization to matrix systems appears possible. An approximate version of the TFD involves coupling the system through single trace interaction \cite{Maldacena:2018lmt, Alet:2020ehp, Plugge:2020wgc}, this corresponds to interaction term in QM that we studied, therefore these models provide a possibility to simulate temperature in the ground state. Likewise will be a fuller exploration of the phase structure of this theory.


\acknowledgments{
The work of RdMK, AJ and JPR is partially funded by a Simons Foundation Grant, Award ID 509116.
The work of RdMK is also supported by the South African Research Chairs initiative 
of the Department of Science and Technology and the National Research Foundation.
}

\appendix

\section{Analytic Results for Matrix Integrals}

For the cases with $k=0$ we can use the techniques \cite{Brezin:1977sv} to get values of loops that the numerics must reproduce. When $k\ne 0$ but $g_3 =g_4 = 0$ we are left with a quadratic integral which is easily performed.

\subsection{Quartic Model}

For the matrix integral with action
\begin{eqnarray}
S = \frac{1}{2}{\rm Tr} M^2 + \frac{g_4}{N}{\rm Tr} M^4 \, .
\end{eqnarray}
The density of eigenvalues obeys
\begin{eqnarray}
-\!\!\!\!\!\!\!\!\;\int_{-2a}^{2a}\,\,\,\frac{\phi (y)}{x-y}\dd y = \frac{1}{x}+2g_4 x^3 \, ,
\qquad\qquad |x|\le 2a \, ,
\end{eqnarray}
and the normalization condition
\begin{eqnarray}
\int_{-2a}^{2a}\phi (x) \dd x =1 \, .
\end{eqnarray}
This is solved by
\begin{eqnarray}
\phi(x)=\frac{1}{\pi}\left(\frac{1}{2}+4g_4a^2+2g_4 x^2\right)\sqrt{4a^2-x^2} \, ,
\end{eqnarray}
where
\begin{eqnarray}
12 g_4 a^4 +a^2 -1 =0 \, .
\end{eqnarray}
Using this density we compute the following planar expectation values
\begin{eqnarray}
\frac{1}{N^2}\langle{\rm Tr} M^2\rangle=\int_{-2a}^{2a}\phi (x) x^2\,\,\dd x
=\frac{\sqrt{48 g_4+1}+24 g_4 \left(2 \sqrt{48 g_4+1}-3\right)-1}{864 g_4^2}
\end{eqnarray}

\begin{eqnarray}
\frac{1}{N^3}\langle{\rm Tr} M^4\rangle =\int_{-2a}^{2a}\phi (x) x^4\,\,\dd x
=\frac{-\sqrt{48 g_4+1}+24 g_4 \left(36 g_4-2 \sqrt{48 g_4+1}+3\right)+1}{3456 g_4^3}\cr
\end{eqnarray}

\begin{align}
\frac{1}{N^4}\langle{\rm Tr} M^6\rangle =&\int_{-2a}^{2a}\phi (x) x^6\,\,\dd x \nonumber \\ 
=&\frac{\sqrt{48 g_4+1}+8 g_4 \left(7 \sqrt{48 g_4+1}+12 g_4 \left(4 \sqrt{48 g_4+1}-15\right)-10\right)-1}{13824 g_4^4}
\end{align}

\begin{align}
\frac{1}{N^5}\langle{\rm Tr} M^8\rangle = & \int_{-2a}^{2a}\phi (x) x^8\,\,\dd x \nonumber \\
 = &-\frac{7}{373248 g_4^5}
     \bigg[\sqrt{48g_4+1}-1 \nonumber \\
   & \quad -6g_4\left(-11 \sqrt{48g_4+1}+72g_4\left(20g_4-2\sqrt{48g_4+1}+5\right)+15\right)\bigg]. 
\end{align}

\subsection{Quadratic Two-Matrix Model}

Doing the Gaussian integrals its simple to find
\begin{align}
\frac{1}{N^2}\langle{\rm Tr} (M_1^2)\rangle  & =   \frac{1}{1-k^2} \qquad
& \frac{1}{N^2}\langle{\rm Tr} (M_1M_2)\rangle & = - \frac{k}{1-k^2} \nonumber \\
\frac{1}{N^3}\langle{\rm Tr} M_1^4\rangle    & = \frac{2}{(1-k^2)^2} \qquad
& \frac{1}{N^3}\langle{\rm Tr} (M_1^2M_2^2)\rangle & = \frac{1+k^2}{1-k^2} \nonumber \\
\frac{1}{N^4}\langle{\rm Tr} (M_1^6)\rangle  & =\frac{5}{(1-k^2)^3} \qquad
& \frac{1}{N^4}\langle{\rm Tr} M_1^2M_2^4\rangle & = \frac{2+3k^2}{(1-k^2)^3} \nonumber \\
\frac{1}{N^4}\langle{\rm Tr} (M_1^4M_2^2)\rangle & =\frac{2+3k^2}{(1-k^2)^3}\qquad
& \frac{1}{N^5}\langle{\rm Tr} (M_1^4M_2^4)\rangle & =\frac{4+9k^2+k^4}{(1-k^2)^4} \nonumber \\
\frac{1}{N^6}\langle{\rm Tr} (M_1^{10})\rangle & =\frac{42}{(1-k^2)^5}\qquad
& \frac{1}{N^6}\langle{\rm Tr} (M_1^4M_2^4)\rangle & = \frac{14+28k^2}{(1-k^2)^5} \nonumber \\
\frac{1}{N^6}\langle{\rm Tr} (M_1^5M_2^5)\rangle & =\frac{25k+16k^3+k^5}{(1-k^2)^5}\qquad
& \frac{1}{N^7}\langle{\rm Tr} (M_1^{12})\rangle & =\frac{132}{(1-k^2)^6} \nonumber \\
\frac{1}{N^7}\langle{\rm Tr} (M_1^6M_2^6)\rangle & =\frac{25+81k^2+25k^4+k^6}{(1-k^2)^6}\qquad
& \frac{1}{N^8}\langle{\rm Tr} (M_1^{14})\rangle & =\frac{429}{(1-k^2)^7} \nonumber \\
\frac{1}{N^8}\langle{\rm Tr} (M_1^{10}M_2^4)\rangle & = \frac{3 \left(25 k^4+90 k^2+28\right)}{(1-k^2)^7}\qquad
& \frac{1}{N^9}\langle{\rm Tr} (M_1^{16})\rangle & = \frac{1430}{(1-k^2)^8}
\end{align}

\subsection{Cubic Model}

For the matrix integral with action
\begin{eqnarray}
S= \frac{1}{2}{\rm Tr} M^2 + \frac{g_3}{\sqrt{N}}{\rm Tr} M^3 \, .
\end{eqnarray}
The density of eigenvalues obeys
\begin{eqnarray}
-\!\!\!\!\!\!\!\!\;\int_{2a}^{2b}\,\,\,\frac{\phi (y)}{x-y}\dd y = \frac{1}{x} + \frac{3}{2}g_3 x^3 \, ,\qquad\qquad 2a\le x\le 2b \, ,
\end{eqnarray}
and the normalization condition
\begin{eqnarray}
\int_{2a}^{2b}\phi (x) \dd x =1 \, .
\end{eqnarray}
This is solved by
\begin{eqnarray}
\phi(x) = \frac{1}{\pi}\left(1+3g_3 (a+b)+3g_3 x\right)\sqrt{(x-2a)(2b-x)} \, ,
\end{eqnarray}
with $a$ and $b$ obtained from
\begin{eqnarray}
3g_3(b-a)^2 + 2(a+b)[1+3g_3(a+b)] =0 \, ,
\end{eqnarray}
\begin{eqnarray}
(b-a)^2[1+6g_3 (a+b)]=4 \, .
\end{eqnarray}
The solutions to these equations are ugly, so we will plug in definite values of $g_3$.
Note that we must take $g_3^2 < \frac{1}{108\sqrt{3}} \approx 0.00534$ to get convergence of planar perturbation theory.
We take $g_3=0.01$ ($a=-1.03198$ and $b=0.971651$),
$g_3=0.025$ ($a=-1.08956$ and $b=0.934167$),
and $g_3=0.01$ ($a=-1.23513$ and $b=0.880559$).
Using this density we compute the following planar expectation values
\begin{eqnarray}
\frac{1}{N^{3/2}}\langle{\rm Tr} M\rangle=\int_{2a}^{2b}\phi (x) x\,\,\dd x
&=&-0.0301088\qquad (g=0.01) \, ,\cr\cr
&=&-0.0767683\qquad (g=0.025) \, ,\cr\cr
&=&-0.166732\qquad (g=0.05) \, .
\end{eqnarray}
\begin{eqnarray}
\frac{1}{N^2}\langle{\rm Tr} M^2\rangle=\int_{2a}^{2b}\phi (x) x^2\,\,\dd x
&=&1.00363\qquad (g=0.01) \, ,\cr\cr
&=&1.02358\qquad (g=0.025) \, ,\cr\cr
&=&1.11155\qquad (g=0.05) \, .
\end{eqnarray}
\begin{eqnarray}
\frac{1}{N^3}\langle{\rm Tr} M^4\rangle =\int_{2a}^{2b}\phi (x) x^4\,\,\dd x
&=&2.02182\qquad (g=0.01) \, , \cr\cr
&=&2.14427\qquad (g=0.025) \, ,\cr\cr
&=&2.73465\qquad (g=0.05) \, .
\end{eqnarray}
\begin{eqnarray}
\frac{1}{N^4}\langle{\rm Tr} M^6\rangle =\int_{2a}^{2b}\phi (x) x^6\,\,\dd x
&=&5.10951\qquad (g=0.01) \, ,\cr\cr
&=&5.73859\qquad (g=0.025) \, ,\cr\cr
&=&9.1135\qquad (g=0.05) \, .
\end{eqnarray}

\section{Analytical Results of The One-Matrix Quantum Mechanics}
\label{appendix:one-matrix-analytical}

We consider the hermitian one-matrix quantum mechanics with a quartic interaction
\begin{equation}
    H = - \frac{1}{2} \Tr(\pdv[2]{}{M}) + \frac{1}{2} \Tr(M^2) + \frac{g_4}{N} \Tr(M^4) \, .
\end{equation}
This model is dual to the $D=1$ string theory in the double scaling limit. The model was first solved in \cite{Kazakov:1988ch} and its spectrum was solved in \cite{Das:1990kaa} using a field theoretic approach. We briefly summarize the analytical results derived there. 

\paragraph{Eigenvalue distribution.}
In the large $N$ limit the eigenvalue distribution of the ground state \cite{Brezin:1977sv, Jevicki:1979mb} is
\begin{equation}
    \phi_0(x) = \frac{1}{\pi}\sqrt{\Lambda^2 + 2 g_4 \Lambda^4 - x^2 - 2g_4 x^4} \, ,
\end{equation}
where $x\in [-\Lambda, \Lambda]$, and $\Lambda$ is determined by the constraint
\begin{equation}
    \int_{-\Lambda}^{\Lambda} \phi_0(x) \mathrm{d}x =  
    \frac{\Lambda^2}{2}\sqrt{1+2 g_4 \Lambda^2} \: {}_{2}F_{1}\left(-\frac{1}{2},\frac{1}{2},2;-\frac{2 g_4 \Lambda^2}{1+2 g_4 \Lambda^2}\right) = 1 \, .
\end{equation}
The constraint is saturated at the critical value $g_{\operatorname{c}} = - 1/ 3 \sqrt{2} \pi$. 

\paragraph{Loop values.}
The loops at the ground state are
\begin{equation}\label{eq:loop-analytical}
\begin{split}
    \phi(2C) \equiv & \frac{\Tr(M^{2C})}{N^{C+1}} 
    = \int_{-\Lambda}^{\Lambda} \phi_0(x) x^{2C} \mathrm{d}x \\
    = &
    \frac{\Lambda^{2C+2} \sqrt{1+2 g_4 \Lambda^2}}{2 \sqrt{\pi}}
    \frac{\Gamma\left(C+\frac{1}{2}\right)}{\Gamma\left(C+2\right)}\: 
    {}_{2}F_{1}\left(-\frac{1}{2},C+\frac{1}{2}, C+2;-\frac{2 g_4 \Lambda^2}{1+2 g_4 \Lambda^2}\right) \, ,
\end{split}
\end{equation}
and all loops with odd powers are zero.

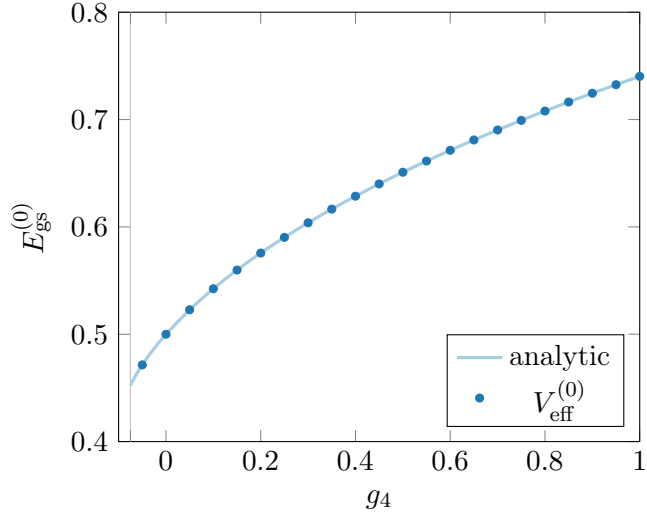
\begin{figure}[htb!]
    \begin{center}
            
\begin{tikzpicture}

\begin{axis}[
    cycle list/Paired-12,
    xmin=-0.1, xmax=1,
    ymin=0.4, ymax=0.8,
    xlabel={$g_4$},
    ylabel={$E_{\operatorname{gs}}^{(0)}$},
    xtick={0,0.2,0.6,0.4,0.8,1},
    ytick={0.4,0.5,0.6,0.7,0.8},
    grid=minor,
    extra x ticks={-0.0750264},
    extra x tick style={grid=major},
    extra x tick labels={},
    legend style={
        at={(0.8,0.25)}, 
        anchor=north,
    },
]

\addplot+[
    smooth,
    mark=,
    line width=1.2pt,
] table {data_1M_ground_state_energy_analytical.dat};
\addlegendentry{analytic}

\addplot+[
    only marks,
    mark=*,
    mark size=1.5pt,
] table {data_1M_ground_state_energy_Veff_value.dat};
\addlegendentry{$V_{\operatorname{eff}}^{(0)}$}

\end{axis}

\end{tikzpicture}
        \caption{One-matrix $E_{\operatorname{gs}}^{(0)}$ versus the quartic coupling $g_4$.}
        \label{fig:1M-ground-state-energy}
    \end{center}
    \end{figure}

\paragraph{Ground state energy.}
The leading ground state energy is
\begin{equation} \label{eq:1M-Egs_0}
    E_{\operatorname{gs}}^{(0)} = 
    \frac{1}{2} \Lambda ^2 \left(1 + 2 g \Lambda ^2\right) 
    -\frac{1}{8} \Lambda ^4 \left(1 + 2 g_4 \Lambda^2\right)^{3/2} \, _2F_1\left(-\frac{3}{2}, \frac{1}{2}, 3; -\frac{2 g_4 \Lambda ^2}{1 + 2 g_4 \Lambda
    ^2}\right) \, .
\end{equation}
In the free case we have $E_{\operatorname{gs}}^{(0)} = 1/2$. Figure \ref{fig:1M-ground-state-energy} shows the results obtained from the above analytic expression, as well as the collective potential $V_{\eff}$ values, demonstrating an excellent agreement.

\paragraph{Spectrum.}
In the large $N$ limit the spectrum of small fluctuations is
\begin{equation}
    \varepsilon_n (g_4) = \frac{n \pi}{2} \left[ \int_0^{\Lambda} \frac{\mathrm{d}x}{\pi \phi_0(x)} \right]^{-1} \, .
\end{equation}
The integral can be be evaluated analytically
\begin{equation} \label{eq:1M_spectrum_analytical}
    \varepsilon_n (g_4) = \frac{n\pi}{2} \: \sqrt{1+2 g_4 \Lambda^2} \: 
    \left[K\left( \sqrt{\frac{- 2 g_4 \Lambda^2}{1+2 g_4 \Lambda^2}}\right)\right]^{-1} \, ,
\end{equation}
where $K$ is the complete elliptic integral of the first kind. The general feature of \eqref{eq:1M_spectrum_analytical} is that the level $n$ frequency is proportional to $n$, no matter how $g_4$ varies. One thus expects that the spectrum fits a straight line when plotted as a function of level number.

\bibliographystyle{jhep}
\bibliography{reference}

\end{document}